\documentclass[11pt]{article}
\usepackage[utf8]{inputenc}
\usepackage[T1]{fontenc}
\usepackage[utf8]{inputenc}
\usepackage{authblk}

\title{\Large{Assortment and Reciprocity Mechanisms for Promotion of Cooperation in a Model of Multilevel Selection}}
\author[1,2]{Daniel B. Cooney}
\affil[1]{Department of Mathematics, University of Pennsylvania, Philadelphia, PA, USA}
\affil[2]{Center for Mathematical Biology, University of Pennsylvania, Philadelphia, PA, USA}
\date{\today}

\usepackage{latexsym}
\usepackage{amssymb,amsmath}
\usepackage{amsthm}
\usepackage{mathtools}
\usepackage{graphicx}
\usepackage{color}
\usepackage{blkarray}
\usepackage{multirow}
\usepackage{hyperref}
\usepackage{float}
\usepackage{subfig}
\usepackage{relsize}
\usepackage{array,makecell}

\usepackage{cleveref}

\usepackage[numbers,sort&compress]{natbib}

\hypersetup{
  colorlinks   = true, 
  urlcolor     = blue, 
  linkcolor    = blue, 
  citecolor   = red 
}

\usepackage{breqn}

\usepackage{xcolor}
\usepackage{cancel}

\usepackage{hyperref}
\usepackage{empheq}

\usepackage[utf8]{inputenc}
\usepackage{nicefrac}

\usepackage{setspace}

\usepackage{enumerate}

\usepackage[titletoc,toc]{appendix}

\usepackage{tocbasic}
\DeclareTOCStyleEntry[
  beforeskip=.2em plus 1pt,
  pagenumberformat=\textbf
]{tocline}{section}

\usepackage[font=small,labelfont=bf]{caption}

\newcommand{\ds}{\displaystyle}

\newcommand{\bbm}{\begin{bmatrix}}
\newcommand{\bpm}{\begin{pmatrix}}
\newcommand{\ebm}{\end{bmatrix}}
\newcommand{\epm}{\end{pmatrix}}

 \newcommand{\dsdel}[2]{\displaystyle\frac{\partial #1}{\partial #2}}

\newcommand{\dsddx}[2]{\displaystyle\frac{d #1}{d #2}}
\newcommand{\dsddt}[1]{\displaystyle\frac{d #1}{dt}}

\renewcommand{\abstractname}{Abstract}
\numberwithin{equation}{section}
\numberwithin{figure}{section}
\renewcommand{\thesection}{\arabic{section}}

\newcommand{\structure}{\varsigma}

\begin{document}

\maketitle

\newtheorem{definition}{Definition}[section]
\newtheorem{theorem}{Theorem}[section]
\newtheorem{lemma}[theorem]{Lemma}
\newtheorem{corollary}[theorem]{Corollary}
\newtheorem{claim}[theorem]{Claim}
\newtheorem{fact}[theorem]{Fact}
\newtheorem{proposition}[theorem]{Proposition}
\newtheorem{remark}[theorem]{Remark}
\newtheorem{example}[theorem]{Example}
\newtheorem{observation}[theorem]{Observation}


\renewcommand{\thesection}{\arabic{section}}
\setcounter{section}{0}

\begin{abstract}
 In the study of the evolution of cooperation, many mechanisms have been proposed to help overcome the self-interested cheating that is individually optimal in the Prisoners' Dilemma game.  These mechanisms include assortative or networked social interactions, other-regarding preferences considering the payoffs of others, reciprocity rules to establish cooperation as a social norm, and multilevel selection involving simultaneous competition between individuals favoring cheaters and competition between groups favoring cooperators. In this paper, we build on recent work studying PDE replicator equations for multilevel selection to understand how within-group mechanisms of assortment, other-regarding preferences, and both direct and indirect reciprocity can help to facilitate cooperation in concert with evolutionary competition between groups. We consider a group-structured population in which interactions between individuals consist of Prisoners' Dilemma games, and study the dynamics of multilevel competition determined by the payoffs individuals receive when interacting according to these within-group mechanisms. We find that the presence of each of these mechanisms acts synergistically with multilevel selection for the promotion of cooperation, decreasing the strength of between-group competition required to sustain long-time cooperation and increasing the collective payoff achieved by the population. However, we find that only other-regarding preferences allows for the achievement of socially optimal collective payoffs for Prisoners' Dilemma games in which average payoff is maximized by an intermediate mix of cooperators and defectors. For the other three mechanisms, the multilevel dynamics remain susceptible to a shadow of lower-level selection, as the collective outcome fails to exceed the payoff of the all-cooperator group.

\end{abstract}

\singlespacing

%
{\hypersetup{linkbordercolor=black, linkcolor = black}
\begin{spacing}{0.01}
\renewcommand{\baselinestretch}{0.1}\normalsize
\tableofcontents
\addtocontents{toc}{\protect\setcounter{tocdepth}{2}}
\end{spacing}
\singlespacing

 \section{Introduction} \label{sec:intro}

 In many biological systems, natural selection operates simultaneously across multiple levels of organization, with tensions arising between evolutonary incentives at different levels. Such conflicts readily arise in a wide variety of settings, including the formation of protocells and the origins of life \cite{hogeweg2003multilevel,szathmary1987group,szathmary1995major}, collective behavior in animal groups \cite{boza2010beneficial}, the evolution of aggressive or cooperative behavior of ant queens \cite{shaffer2016foundress}, host-microbe mutualisms in the microbiome \cite{van2019role}, and competition between pathogen strains under both immunological dynamics and epidemiological dynamics \cite{gilchrist2004optimizing,levin1981selection,blackstone2020variation}. Across these various systems, a common theme that arises is an evolutionary tug-of-war between individual-level competition favoring cheaters and higher-level competition favoring groups of cooperators. The evolution of cooperation provides a useful case study for questions of multilevel selection, and evolutionary game theory provides an instructive analytical framework for analyzing the tension between the interests of a group and the interests of the individuals comprising the group 
\cite{traulsen2005stochastic,traulsen2006evolution,traulsen2008analytical,simon2010dynamical,markvoort2014computer,bottcher2016promotion}.

 In the literature on the evolution of cooperation, there has been an emphasis on the role of mechanisms that can help to facilitate the possibility of promoting cooperation when cheating behaviors dominate under individual-level selection in a well-mixed population. Examples of commonly studied mechanisms include assortative interactions \cite{grafen1979hawk,eshel1982assortment}, other-regarding preference \cite{smith1982evolution}, the punishment of defectors via both reciprocal altruism (direct reciprocity) \cite{trivers1971evolution} and social reputations (indirect reciprocity) \cite{nowak2005indirect,ohtsuki2006leading}, and embedding evolutionary game dynamics in space or on social networks \cite{durrett1994importance,killingback1996spatial,ohtsuki2006simple,ohtsuki2006replicator}.
These mechanisms have been shown to facilitate cooperation in infinite populations via individual-level selection using a replicator equation approach, showing that the mechanisms can either promote long-time equilibrium coexistence of cooperators and defectors or the stabilization of the all-cooperator equilibrium for populations featuring a sufficient initial cohort of cooperators. %
These within-group mechanisms have often been compared with multilevel selection as alternative approaches for showing how to promote collectively beneficial cooperation over the individual temptation to defect \cite{nowak2006five,taylor2007transforming}. However, the models of multilevel selection considered in these comparisons rely on stochastic framework modeling individual-level and group-level events in a finite population \cite{traulsen2005stochastic,traulsen2006evolution,traulsen2008analytical}, differing from the replicator equation approach for studying the mechanisms that operate via individual-level selection. 

However, recent work on multilevel selection has introduced and analyzed PDE analogues of replicator equations to study evolutionary dynamics operating at two competing levels. This framework provides a new opportunity for comparing and studying the combined effects of within-group mechanisms of assortment and reciprocity with multilevel selection in the promotion of cooperation in evolutionary games. Luo first introduced a stochastic description for a two-level birth-death process in a group-structured population with two types of individuals, one with an advantage under individual-level reproduction and the other conferring an advantage to its group in group-level reproduction. Considering the limit of infinite group size and infinitely many groups, Luo and coauthors derived a hyperbolic PDE describing the evolutionary competition at both levels, and characterized the long-time behavior of the PDE to determine the long-time support for cooperative behaviors \cite{luo2014unifying,van2014simple,luo2017scaling}. This model was further generalized to study multilevel competition when individual-level and group-level reproduction rates depend on the payoffs from two-strategy social dilemmas \cite{cooney2019replicator,cooney2019analysis}, and to model a broad class of models with continuously differentiable reproduction rates \cite{cooney2021long}. This PDE framework provides an analytically tractable approach to study the conflicts between these levels of selection, and can be extended to include more realistic aspects of between-group competition by incorporating models of group-level fission events \cite{simon2010dynamical,simon2012numerical,simon2013towards,simon2016group,puhalskii2017large}.  

For these PDE models of multilevel selection, there exists a threshold relative strength of selection at the two levels such that cooperation can survive in the long-time population when between-group competition is sufficiently strong \cite{luo2017scaling,cooney2019replicator,cooney2019analysis,cooney2021long}. In such cases, the long-time population can sustain a range of levels of cooperation in the population at a density steady state, which results from the balancing of within-group competition favoring defectors and between-group favoring groups with cooperators. A striking behavior seen in these models is that within-group competition casts a long shadow on the multilevel dynamics: for games in which groups are best off with a mix of cooperators and defectors, no level of between-group competition strength can result in collective payoffs exceeding that of the all-cooperator group \cite{cooney2019replicator,cooney2019analysis,cooney2021long}. For such games, the level of cooperation achieved by the multilevel dynamics will always be less than the composition which maximizes the average payoff of group members, even in the limit of infinitely strong between-group competition. Given the limitations on the ability to promote optimal collective benefits via multilevel selection, from the possibility of insufficient strength of between-group competition to promote cooperation to the shadow of lower-level selection, it is natural to ask whether the addition of within-group mechanisms can further facilitate the evolution of cooperation in concert with multilevel selection.

In this paper, we study the role of the within-group mechanisms of assortment, other-regarding preferences, and both direct and indirect reciprocity on the dynamics of the evolution of cooperation via multilevel selection. We do this by incorporating models for these mechanisms into the PDE replicator equation for multilevel selection, exploring how the modified payoffs generated by each mechanism impact the individual-level incentive to defect and the group-level incentive to cooperate in our model of multilevel selection. For each mechanism we consier, we find that the presence of the mechanism decreases the threshold strength of between-group competition needed to sustain long-time cooperation relative to the case of multilevel selection based upon well-mixed within-group interactions, and similarly we find that each mechanism increases the long-time average payoff achieved for a given strength of between-group competition. We see a difference, however, when comparing how the mechanisms impact the collective outcome in the limit of strong between-group competition. Under the mechanisms of assortment, direct reciprocity, and indirect reciprocity, the multilevel dynamics will always produce as much cooperation as possible in the limit of strong between-group competition, even for games in which an intermediate level of cooperation is optimal for the group. By contrast, the model of other-regarding preference always produces the socially optimal level of cooperation for the limit in which individuals care equally about their own payoff and the payoff of their opponents. While all four mechanisms help faciliate the evolution of cooperation via multilevel selection, our model of other-regarding preference is the only mechanism we consider that helps to erase the shadow of lower-level selection. 

The decrease we see in the threshold relative selection strength required to sustain long-time cooperation highlights the synergistic effects between incorporating competition between groups and within-group population structure on the outcome of evolutionary dynamics across scales. In particular, we find that there are parameter regimes in which neither our within-group mechanism nor multilevel selection cannot produce any cooperation on their own, while the combination of the two mechanisms can allow for long-time cooperation. Multilevel selection has also been attributed as a means by which within-group mechanisms can evolve \cite{boyd2003evolution,santos2007multi,janssen2014effect}, as competition between groups can select for groups that have established mechanisms that are conducive to supporting cooperative traits or behaviors. As a result, these synergistic effects between multilevel competition and within-group mechanisms can be seen playing an important role in the evolution of cooperative groups and the emergence of new evolutionary individuals operating at a higher level of selection \cite{szathmary1995major,michod1996cooperation,michod1997cooperation,nowak2006five}. From a biological perspective, we can view this exploration of within-group modifications as serving as an initial step towards trying to use PDE models of multilevel selection to understand major evolutionary transitions \cite{szathmary1995major}. In a forthcoming paper, a first attempt at exploring such an evolutionary transition has been made for the case of protocell evolution and the origin of chromosomes \cite{gabriel1960primitive,smith1993origin,szathmary1993evolution}, in which a PDE model shows how modification of gene-level and cell-level replication rates can help to overcome the shadow of lower-level selection \cite{cooney2021pde}. We hope this approach for studying PDE models of multilevel selection with modified within-group interactions can be further applied as a tractable way to illustrate how evolutionary transitions can arise in a range of biological and cultural systems.

 The remainder of the paper is structured as follows. In Section \ref{sec:gamesandgeneral}, we present the original model for the multilevel replicator dynamics, and describe existing results for PDE models of multilevel selection that we will apply when analyzing the effects of each of our within-group mechanisms. In Section \ref{sec:assortment}, we present our results for the multilevel dynamics incorporating  the within-group mechanism of like-with-like assortment and, in Section \ref{sec:kinother}, we present results for the multilevel model incorporating other-regarding preferences. In Section \ref{sec:reciprocity}, we present our analysis of the multilevel dynamics for the models of reciprocity, covering indirect reciprocity in Section \ref{sec:indirectreciprocity} and direct reciprocity in Section \ref{sec:directreciprocity}. We conclude in Section \ref{sec:discussion} with a recap of the behaviors found for multilevel selection across our range of mechanisms, and discuss implications for further work on the impact of changing within-group interactions on the evolution of cooperation via multilevel selection.

\section{PDE Model for Multilevel Selection in Evolutionary Games} \label{sec:gamesandgeneral}

In this section, we provide the necessary background on the PDE replicator equation for describing multilevel selection when within-group and between-group competition depends on payoffs obtained by playing two-player, two-strategy social dilemmas. In Section \ref{sec:pdemodel}, we present the payoff functions for these games and show how to use these payoffs to formulate our baseline PDE model. In Section \ref{sec:existingresults}, we present existing results for the long-time behavior for our baseline PDE model, which we will apply in subsequent sections to explore the impact of within-group mechanisms on the dynamics of multilevel selection.

\subsection{PDE Replicator Equation for Two-Strategy Social Dilemmas}
\label{sec:pdemodel}

Here, we will illustrate the baseline model for deterministic multilevel selection for evolutionary games with well-mixed strategic interactions, as introduced in previous work \cite{cooney2019replicator}. We consider two-strategy games, in which individuals can either choose to Cooperate ($C$) or Defect ($D$), and individuals receive payoff from pairwise interaction given by the following payoff matrix

\begin{equation} \label{eq:payoffmatrix}
\begin{blockarray}{ccc}
& C & D \\
\begin{block}{c(cc)}
C & R & S \\
D & T & P \\
\end{block}
\end{blockarray}
\end{equation}
where the payoffs parameters are respectively named ${\bf R}$eward, ${\bf S}$ucker,  ${\bf T}$emptation, and ${\bf P}$unishment. Four games of interest to our analysis are the Prisoners' Dilemma (PD), the Hawk-Dove (HD) game, the stag hunt (SH), and the Prisoners' Delight (PDel), which are characterized by the following rankings of payoffs

\begin{subequations} \label{eq:gamerankings} \begin{align} \mathrm{PD}  :& \: T > R > P > S  \label{eq:PDranking} \\ \mathrm{HD} :&  \: T > R > S > P \\ \mathrm{SH} :& \: R > T > P > S \\ \mathrm{PDel} :& \; R > T > S > P \end{align} \end{subequations}
 \cite{nowak2006evolutionary}. These four games serve as examples of social dilemmas for which competition between individuals favors four different possible long-time outcomes: the PD promotes dominance of defectors, the HD game promotes coexistence of cooperators and defectors, the SH produces bistability between dominance of defectors and dominance of cooperators, and the PDel promotes dominance of cooperators. For each of these games $R > P$, so a group composed entirely of cooperators receives a higher average payoff than a group composed entirely of defectors. 

Four other two-player, two-strategy social dilemmas that have been studied in the context of multilevel selection are characterized by the following rankings of payoffs:
\begin{subequations} \label{eq:alternategames}
\begin{align}
 \textnormal{coordination game 1 (CG1)}  :& \: R > P > S > T  \\ 
 \textnormal{coordination game 2 (CG2)} :& \: R > P > T > P \\
 \textnormal{anti-coordination game 1 (AC1)} :& \:  T > S > R > P   \\ 
\textnormal{anti-coordination game 2 (AC2)} :&  \: S > T > R > P 
 \end{align}
\end{subequations}
\cite{cooney2019analysis}. The two coordination games (CG1,CG2) are similar to the SH game in that they promote bistability between dominance of defectors and dominance of cooperators, while the anti-coordination game (AC1,AC2) are similar to the HD game in that they promote coexistence of cooperators and defectors under individual-level selection. In this paper, we will primarily focus on interactions between individuals that consist of PD games, but our incorporation of mechanisms of assortment, other-regarding preferences, and reciprocity will produce scenarios such that the effective payoffs that shape the birth rates of cooperators and defectors that correspond to each of the eight games introduced above.

 In a group composed of fraction $x$ cooperators and $1-x$ defectors, the expected payoffs received by cooperators and defectors in well-mixed interactions are \begin{subequations} \label{eq:piofx} \begin{align}  
\pi_C(x) &= xR + (1-x) S \\
\pi_D(x) &= x T + (1-x) P
\end{align}
\end{subequations}
and the average payoff of individuals in a group with $x$ fraction cooperators is 
\begin{equation}  \label{eq:Gofx} G(x) = x \pi_C(x) + (1-x) \pi_D(x) = P +  \left( S + T - 2P \right)  x + \left( R - S - T + P \right) x^2  \end{equation}

Using the framework of Luo and Mattingly for nested birth-death processes \cite{luo2014unifying,luo2017scaling}, we can describe the dynamics of multilevel selection in a group-structured population in which within-group competition follows individual payoff and between-group competition depends on the average payoff of group members. In a group composed of a fractions $x$ cooperators and $1-x$ defectors, we assume that cooperators and defectors reproduce at rate $1 + w_I \pi_C(x)$ and $1 + w_I \pi_D(x)$, respectively, and the offspring individuals replace a randomly chosen member of the group. To model between-group competition, we assume that groups featuring a fraction $x$ of cooperators produce copies of themselves at rate $\Lambda \left(1 + w_G G(x)\right)$, replacing a randomly chosen group in the population. Here, the parameters $w_I$ and $w_G$ describe the respective importance of individual-level and group-level payoff on the rate of reproduction events, and $\Lambda$ describes the relative rate of within-group and between-group selection events. 

In the limit in which there are infinitely many groups and each group has infinite size, we can describe the composition of strategies in our group-structured population by the probability density $f(t,x)$, characterizing the density of $x$-cooperator groups at time $t$. It has been shown in prior work that, in this limit, the dynamics of the two-level birth-death process described above can will evolve according to the following partial differential equation for the density $f(t,x)$:
\begin{dmath} \label{eq:pdereplicator} %
\dsdel{f(t,x)}{t} =  -  \overbrace{  \dsdel{}{x} \left( x(1-x) ( \pi_C(x) - \pi_D(x) )  f(t,x) \right)}^{\text{Within-Group Competition}}  + \lambda \underbrace{ f(t,x) \left[ G(x)  - \int_0^1 G(y) f(t,y) dy  \right]}_{\text{Between-Group Competition}},
 \end{dmath}
where the parameter $\lambda := \frac{\Lambda w_G}{w_I}$ measures the relative strength of within-group and between-group competition. The dynamics of multilevel selection provided by Equation \eqref{eq:pdereplicator} must be supplied with an initial density $f(0,x) := f_0(x)$ describing the composition of strategies within groups when $t=0$.

Equation \eqref{eq:pdereplicator} is a first-order hyperbolic PDE, whose time-dependent solutions can be analyzed using the method of characteristics \cite{strauss2007partial,evans1998partial}. The advection term describes the effect of within-group competition. The characteristic curves  are determined by the following ODE
\begin{equation} \label{eq:withincharacteristics}
\begin{aligned}
\dsddx{x(t)}{t} &= x (1-x) \left( \pi_C(x) - \pi_D(x) \right) \\  x(0) &= x_0, 
\end{aligned}
\end{equation}
which is the replicator equation describing individual-level selection within groups \cite{taylor1978evolutionary,hofbauer1998evolutionary,sandholm2010population,cressman2014replicator}. The nonlocal nonlinear term $\lambda f(t,x) \left[G(x) - \int_0^1 G(y) f(t,y) dy \right]$ describes the impact of between-group competition, and resembles an integro-differential replicator equation for group selection. Taking the two terms together, we can think of Equation \eqref{eq:pdereplicator} as the replicator equation describing the simultaneous effect of selection operating at two levels. 

We can also use the expressions for $\pi_C(x)$, $\pi_D(x)$, and $G(x)$ from Equations \eqref{eq:piofx} and \eqref{eq:Gofx} to understand how the the entries of the payoff matrix of Equation \eqref{eq:payoffmatrix} impact the dynamics of multilevel selection. Using the shorthand notation, $\alpha = R - S - T + P$, $\beta = S - P$, and $\gamma = S + T - 2P$, the multilevel dynamics can be rewritten as
\begin{equation} \label{eq:pdeparam}
 \dsdel{f(t,x)}{t} = - \dsdel{}{x} \left[ x(1-x) (\beta + \alpha x)  f(t,x) \right] + \lambda f(t,x) \left[ \gamma x + \alpha x^2  -  \left( \gamma M_1^f(t) + \alpha M_2^f(t) \right) \right],
\end{equation}
where $M_j^f(t) :=  \int_0^1 x^j f(t,y) dy$ denotes the moments of the density $f(t,x)$. In this notation, the average payoff of group members takes the form
\begin{equation}
G(x) = P + \gamma x + \alpha x^2.
\end{equation}
Average payoff $G(x)$ is maximized by the fraction $x^*$ of cooperators given by
\begin{equation} \label{eq:interiormax}
x^* =  \left\{
     \begin{array}{cr}
      \ds\frac{\gamma}{-2\alpha} & : \gamma + 2 \alpha < 0\\
       1 & : \gamma + 2 \alpha \geq 0
     \end{array}
   \right.
   = 
   \left\{
     \begin{array}{cr}
       \ds\frac{S+T-2P}{2 \left(S + T - R - P \right)} & : 2R  < S + T\\
       1 & :  2R \geq S + T
     \end{array}
   \right. 
\end{equation}
\cite{cooney2019replicator,cooney2019analysis}, so the collective outcome of a group can be most favored by a mix of cooperators and defectors when the interaction between a cooperator and defector produces a better contribution to average payoff ($S+T$) than produced by an interaction between two cooperators ($2R$). 

Using this parametrization, the within-group dynamics take the form
\begin{equation}
\dsddt{x(t)} = x (1-x) \left( \beta + \alpha x \right),
\end{equation}
whose equilibria are the all-defector group $x=0$, the all-cooperator group $x=1$, and a third equilibrium $x_{eq}$ given by
\begin{equation}
x_{eq} = \frac{\beta}{-\alpha} = \frac{S - P}{S - P + T - R}.
\end{equation}
The equilibrium $x_{eq}$ is only biologically feasible for the HD, SH, coordination, and anti-coordination games. For two-strategy social dilemmas, there are four generic behaviors of interest for the within-group dynamics. For the PD game, the full-defector equilibrium is globally stable under the within-group replicator dynamics, while the PDel game features global stability of the all-cooperator equilbrium. For the HD and anti-coordination games (AC1,AC2), the interior equilibrium $x_{eq}$ is globally stable, resulting in long-time coexistence of cooperators and defectors. For the SH game and the coordination games (CG1,CG2), the all-cooperator and all-defector equilibria are both locally stable, with basins of attraction that are separated by the equilibrium $x_{eq}$. 

\begin{remark} \label{rem:PDparameters}
For PD games, we can use the payoff ranking from Equation \eqref{eq:PDranking} to see that $\beta = S - P < 0$. However, the parameters $\alpha$ and $\gamma$ can take either sign within the class of PD games. From Equation \eqref{eq:PDranking}, we do have the constraint that $\gamma + \alpha = (S+T-2P) + (R - S - T + P) = R - P > 0$, which means that at least one of $\alpha$ or $\gamma$ must be positive for any PD game. When, in subsequent sections, we consider the impact of various mechanisms on the payoffs received from PD games, we will therefore need to consider underlying PD games with payoff matrices satisfying either $\gamma, \alpha > 0$, $\gamma < 0$ and $\alpha > 0$, or $\gamma > 0$ and $\alpha < 0$. For the first two cases, the average payoff $G(x)$ under well-mixed interactions is always maximized by the all-cooperator composition, while an intermediate level of cooperation can maximize $G(x)$ for the third case (in which $\alpha < 0$) under the additional assumption that $\gamma + 2 \alpha < 0$ (as described in Equation \eqref{eq:interiormax}). 
\end{remark}

\subsection{Existing Results on the Long-Time Behavior of PDE Models for Multilevel Selection}
\label{sec:existingresults}

In this section, we will highlight existing results for the long-time behavior for solutions to the multilevel dynamics of Equation \eqref{eq:pdereplicator}, explaining how the long-time distribution of cooperation depends on the payoff matrix of the underlying game, the relative strength of between-group competition $\lambda$, and the initial distribution of strategies $f_0(x)$. We first present the results for the range of possible two-player, two-strategy social dilemmas presented above, and then, in Section \ref{sec:transformedpayoff} show how these results may be applied to analyze the mechanisms of assortment and reciprocity that we consider in this paper.

In prior work on PDE models of multilevel selection, it has been shown that the long-time behavior of $f(t,x)$ depends on a a property of the tail of the initial strategy distribution known as the H{\"o}lder exponent near $x=1$ \cite{luo2017scaling,cooney2019replicator,cooney2019analysis,cooney2021long}. This quantity measures how the probability contained in the interval $[1-y,1]$ behaves as as $y \to 0$, and has the following formal definition. 
\begin{definition} \label{def:Holder} A probability distribution with density $f(t,x)$ has a H{\"o}lder exponent $\theta_t$ with associated H{\"o}lder constant $C_{\theta_t}$ near $x=1$ if the density has the following limiting behavior %
\begin{equation} \label{eq:Holderlimit}
\lim_{x \to 0} \frac{\int_{1-x}^1 f(t,y) dy}{x^{\Theta}} = \left\{
     \begin{array}{lr}
       0 & : \Theta < \theta_t \\
       C_{\theta_t} & : \Theta =  \theta_t \\
       \infty & : \Theta > \theta_t
     \end{array}
   \right. .
\end{equation}
 \end{definition}
 By plugging into Equation \eqref{eq:Holderlimit}, we see that the family of densities $f^{\theta}(x) = \theta (1-x)^{\theta - 1}$ provide examples of probability densities with H{\"o}lder exponent $\theta$ near $x=1$, each with associated H{\"o}lder constant $C_{\theta} = 1$. As a result, we can think of the H{\"o}lder exponent $\theta$ for a density as describing like which power $(1-x)^{\theta -1}$ does probability of the density $f(t,x)$ vanishs near $x=1$. For increasing H{\"o}lder exponent $\theta$ of our initial density $f_{0}(x)$, the initial population will feature a decreasing portion of groups with compositions close to the all-cooperator equilibrium. 
 
 The H{\"o}lder exponent near $x=1$ has been shown to be preserved by the multilevel dynamics of Equation \eqref{eq:pdereplicator} \cite{cooney2019analysis}. In addition, there is at most one steady state density for the multilevel dynamics with a given H{\"o}lder exponent $\theta$ \cite{cooney2019analysis,cooney2021long}, highlighting the key role played by the H{\"o}lder exponent of the initial population in characterizing the possible long-time behavior of solutions to Equation \eqref{eq:pdereplicator}. For the PD game, the unique steady state density with H{\"o}lder exponent $\theta > 0$ near $x=1$ takes the following form
 \begin{equation}\label{eq:PDsteady}
f^{\lambda}_{\theta}(x) = Z_f^{-1} x^{|\beta|^{-1} \left[ \lambda \left( \gamma + \alpha\right) - \left(|\beta| - \alpha\right) \theta \right]- 1}(1-x)^{\theta-1} \left( |\beta| - \alpha x \right)^{-\frac{\lambda}{|\beta|} \left( \gamma + \alpha + |\beta| \right) - \frac{\alpha}{|\beta|} \theta - 1},
\end{equation}
where $Z_f^{-1}$ is a normalizing constant. This density will only be integrable near $x=0$ if the exponent of $x$ exceeds $-1$, which occurs when that the between-group selection strength satisfies
 \begin{equation} \label{eq:PDthresh} \lambda > \lambda^*_{PD} := \frac{-\left( \beta + \alpha \right)\theta}{\gamma + \alpha} = \frac{\left( \pi_D(1) - \pi_C(1) \right) \theta}{G(1) - G(0) }. \end{equation}
 This threshold selection strength illustrates how the survival of long-time cooperation requires success of many-cooperator groups in the tug-of-war between the individual incentive to defect in compositions near the all cooperator equilibrium $\pi_D(1) - \pi_C(1)$ and the collective incentive to achieve full-cooperation rather than full-defection $G(1) - G(0)$. 
 For HD games, the steady state density with H{\"o}lder exponent $\theta > 0$ near $x=1$ takes the form
 \begin{equation} \label{eq:HDsteady}
g^{\lambda}_{\theta}(x) =  \left\{
     \begin{array}{cr}
       0 &: x < \frac{\beta}{|\alpha|} \\
    Z_g^{-1} x^{\beta^{-1} \left[ \lambda \left(|\alpha| - \gamma\right) + \left(|\alpha| - \beta \right) \theta \right] - 1} \left( 1 - x \right)^{\theta - 1} \left(|\alpha| x - \beta\right)^{\beta^{-1} \left[ \lambda \left(\gamma - |\alpha| - \beta \right) - |\alpha| \theta \right] - 1} &: x \geq \frac{\beta}{|\alpha|}   \end{array}
   \right.,
\end{equation}
 supporting levels of cooperation between the within-group equilibrium $x_{eq} = \frac{\beta}{-\alpha}$ and the all-cooperator group $x=1$. This density will be integrable near $x = x_{eq} = \frac{\beta}{-\alpha}$ when the exponent of $|\alpha| x - \beta$ exceeds $-1$, which occurs provided that between-group selection strength satisfies
\begin{equation} \label{eq:HDthresh} \lambda >  \lambda^*_{HD} := \frac{-\alpha \theta}{\gamma + \alpha - \beta}= \frac{\left( \pi_D(1) - \pi_C(1) \right) \theta}{G(1) - G(x_{eq}) }. \end{equation}
For PD games when $\lambda < \lambda^*_{PD}$, HD games when $\lambda < \lambda^*_{HD}$, and all of the other games mentioned above, the only other possible steady state solutions to Equation \eqref{eq:pdereplicator} are delta-functions concentrated at equilibria of the within-group dynamics. 

In recent work, the long-time behavior of solutions to the multilevel dynamics have been characterized for a class of models including the games with the payoff rankings presented above \cite{cooney2019analysis,cooney2021long}. In Theorem \ref{thm:longtimesummary}, we summarize results for convergence to steady state across the range of two-player, two-strategy social dilemmas given an initial population with density $f_0(x)$ with H{\"o}lder exponent $\theta > 0$ near $x=1$. We characterize the long-time behavior using the notion of weak convergence. In particular, for a limit function $f_{\infty}(x)$, we say that $f(t,x) \rightharpoonup f_{\infty}(x)$ ($f(t,x)$ converges weakly to $f_{\infty}(x)$) if, for all continuous function $v(x)$, $\int_0^1 v(x) f(t,x) dx \to \int_0^1 v(x) f_{\infty}(x) dx$.  %

\begin{theorem} \label{thm:longtimesummary}
Suppose the initial population is a probability density $f(0,x) = f_{\theta}(x)$ with H{\"o}lder exponent $\theta > 0$ and finite H{\"o}lder constant $C_{\theta} > 0$ near $x=1$. 
\begin{itemize}
\item For PD games, the population converges to a delta-function concentrated at the all-defector equilibrium $f(t,x) \rightharpoonup \delta(x)$ as $t \to \infty$ if $\lambda \leq \lambda^*_{PD}$ \cite[Theorem 1.8 and Proposition 5.3]{cooney2021long}. If $\lambda > \lambda^*_{PD}$, the population converges to the unique density steady state with H{\"o}lder exponent $\theta > 0$ near $x=1$: $f(t,x) \rightharpoonup f^{\lambda}_{\theta}(x)$ where $f^{\lambda}_{\theta}(x)$ is given by Equation \eqref{eq:PDsteady} \cite[Theorem 1.2]{cooney2021long}.
\item For HD games, the population converges to a delta-function concentrated at the within-group HD equilibrium $f(t,x) \rightharpoonup \delta(x-x_{eq})$ as $t \to \infty$ if $\lambda \leq \lambda^*_{HD}$ \cite[Theorem C.2]{cooney2021long}. If $\lambda > \lambda^*_{HD}$, the population converges to the unique density steady state with H{\"o}lder exponent $\theta > 0$ near $x=1$: $f(t,x) \rightharpoonup g^{\lambda}_{\theta}(x)$ where $g^{\lambda}_{\theta}(x)$ is given by Equation \eqref{eq:HDsteady} \cite[Theorem C.4]{cooney2021long}.
\item For all games in which the all-cooperator equilibrium is locally stable (PDel, SH, CG1, CG2), the population converges to a delta-function at the all-cooperator equilibrium, $f(t,x) \rightharpoonup \delta(x-1)$ as $t \to \infty$, whenever there is any between-group competition (i.e. $\lambda > 0$) \cite[Proposition 8]{cooney2019analysis}.
\item For the anti-coordination games (AC1,AC2), the population converges to the within-group equilibrium, $f(t,x) \rightharpoonup \delta(x-x_{eq})$ as $t \to \infty$, for any level of between-group competition $\lambda \geq 0$ \cite[Proposition 9]{cooney2019analysis}. 
\end{itemize}
\end{theorem}
While the HD and anti-coordination games (AC1,AC2) both feature within-group dynamics with a globally stable within-group equilibrium $x_{eq}$, the two types of games differ in the landscape they generate for between-group competition. For the HD game, the all-cooperator equilibrium features a higher average payoff from group members than that of the within-group equilibrium $x_{eq}$ ($G(1) > G(x_{eq})$), while the opposite is true for the anti-coordination games ($G(1) < G(x_{eq})$). As a result, both within-group and between-group competition promotes movement toward the within-group equilibrium for the anti-coordination games, yielding long-time concentration upon a delta-function the within-group equilibrium $\delta(x-x_{eq})$ for any strength of between-group competition. 
 
For the case of PD and HD games, we can also characterize useful properties of the steady state densities to quantify the collective outcome from maintaining cooperation in the long-time limit. For the PD steady states, we can calculate that the average payoff of the population at density steady state is given by
\begin{equation} \label{eq:steadystatefitness} 
\begin{aligned} \langle G(\cdot) \rangle_{f^{\lambda}_{\theta}} := \int_0^1 G(x) f^{\lambda}_{\theta}(x) dx  &= P +  \gamma + \alpha + \frac{\left(\beta + \alpha\right) \theta}{\lambda} \\  &=  G(1) - \frac{\left(\pi_D(1) - \pi_C(1)\right) \theta}{\lambda} 
\end{aligned}
\end{equation}
for $\lambda \geq \lambda^*_{PD}$ \cite{cooney2019analysis}. In addition, we can use Equation \eqref{eq:PDthresh} to write the average payoff of the population at the density steady state using the threshold selection intensity $\lambda^*_{PD}$, which gives us
\begin{equation} \label{eq:steadystatefitnesslambdastar} \langle G(\cdot) \rangle_{f^{\lambda}_{\theta}} = \left( \frac{\lambda^*_{PD}}{\lambda}\right) G(0) + \left( 1 - \frac{\lambda^*_{PD}}{\lambda} \right) G(1) \end{equation}
\cite{cooney2019analysis,cooney2021long}. Therefore the average payoff of the population at steady state interpolates between $G(0)$ at $\lambda = \lambda^*_{PD}$ and $G(1)$ as $\lambda \to \infty$. Notably, this means that the average payoff of the population is limited by the average payoff of full cooperator group $G(1)$, even for cases of the PD game in which the average payoff of group members $G(x)$ is maximized by an intermediate fraction of cooperators and defectors. An analogous behavior is seen for HD games when $\lambda \geq \lambda^*_{HD}$, where the average payoff at steady state satisfies
\begin{equation}
\langle G(\cdot) \rangle_{g^{\lambda}_{\theta}} =  \left( \frac{\lambda^*_{HD}}{\lambda}\right) G(x_{eq}) + \left( 1 - \frac{\lambda^*_{HD}}{\lambda} \right) G(1)
\end{equation}
\cite{cooney2019analysis}. From these expressions for the collective payoff at steady state, we see a signature of a shadow of lower-level selection: the presence of even a slight individual-level advantage for defection prevents the long-time achievement of intermediate collective payoff optima under the multilevel dynamics of Equation \eqref{eq:pdereplicator}. 

We can also explore this shadow of lower-level selection by computing the modal fraction of cooperation in a given steady state density. For both the PD and HD game, we can compute that, when $\lambda$ is large enough such that the multilevel dynamics converge to a bounded density state, the modal level of cooperation in the long-time population is given by the following expression:
\begin{equation} \label{eq:modalshared}
\hat{x}_{\lambda}^{-} = \frac{\lambda \gamma - 2 (\alpha -\beta) - \ds\sqrt{\left(\lambda \gamma - 2 (\alpha -\beta) \right)^2 + 4 \left(\lambda (\gamma + \alpha) + (\beta + \alpha) \theta + \beta \right) ( 3 + \lambda) \alpha}}{-2 ( 3 + \lambda)\alpha}
\end{equation}
\cite{cooney2019analysis}. For the PD case, we further note that $f(t,x)$ converges to $\delta(x)$ when $\lambda \leq \lambda^*_{PD}$ and that the steady state density of Equation \eqref{eq:PDsteady} blows up near $0$ when $\lambda^*_{PD} < \lambda < \lambda^*_{PD} + \frac{|\beta|}{\gamma + \alpha}$. This tells us that the modal outcome at steady state for the PD game satisfies the following piecewise characterization
\begin{equation} \label{eq:PDmodal}
\hat{x}_{\lambda} =  \left\{
     \begin{array}{lr}
       0 & : \lambda \leq \lambda^*_{PD} + \frac{|\beta|}{\gamma + \alpha} \\
       \hat{x}_{\lambda}^{-} & :  \lambda > \lambda^*_{PD} + \frac{|\beta|}{\gamma + \alpha} 
     \end{array}
   \right. 
\end{equation}
\cite{cooney2019analysis}. Using an analogous approach, we can find that the modal composition at steady state for the multilevel HD dynamics is given by 
\begin{equation} \label{eq:HDmodal}
\hat{x}_{\lambda} =  \left\{
     \begin{array}{cr}
       x_{eq} & : \lambda \leq \lambda^*_{HD} + \frac{\beta}{\gamma + \alpha - \beta} \\
       \hat{x}_{\lambda}^{-} & :  \lambda > \lambda^*_{HD} + \frac{\beta}{\gamma + \alpha - \beta} 
     \end{array}
   \right. 
\end{equation}
\cite{cooney2019analysis}. In the limit of infinite strength of between-group competition, we can further use Equation \eqref{eq:modalshared} to compute that the modal composition at steady state satisfies
\begin{equation} \label{eq:modallimit}
\hat{x}_{\infty} := \lim_{\lambda \to \infty} \hat{x}_{\lambda} =   \left\{
     \begin{array}{lr}
      \ds\frac{\gamma + \alpha}{-\alpha} & : \gamma + 2 \alpha < 0\\
       1 & : \gamma + 2 \alpha \geq 0
     \end{array}
   \right.
\end{equation}
\cite{cooney2019analysis}. For the case of PD and HD games for which there is an intermediate collective payoff optimum $x^* = \frac{\gamma}{-2\alpha} < 1$ (and correspondingly $\gamma + 2 \alpha < 0$), this means that $\hat{x}_{\infty} = \lim_{\lambda \to \infty} \hat{x}_{\lambda} = \frac{\gamma + \alpha}{-\alpha} < \frac{\gamma}{-2\alpha} = x^*$ \cite{cooney2019analysis}. As a result, the population will concentrate upon a composition feature less cooperation than is socially optimal even when we our relative emphasis $\lambda$ on between-group composition tends to infinity \cite{cooney2021long}. In the alternate case in which $G(x)$ is maximized by the all-cooperator outcome $x^* = 1$, we instead have that the modal composition satisfies $\hat{x}_{\infty} = 1 = x^*$, and the population concentrates upon the socially optimal all-cooperator composition in the limit as $\lambda \to \infty$ \cite{cooney2019analysis,cooney2021long}. 

In the remainder of our paper, we will use these expressions for the average payoff and modal composition of cooperation at steady state to quantify the impact of within-group mechanisms on supporting long-time cooperation. We will place emphasis on how these mechanisms impact the ability to faciliate cooperation at lower relative strengths of between-group selection $\lambda$, and we will explore the behavior of the model in the limit as $\lambda \to \infty$ to see how the mechanisms interact with the shadow of lower-level selection. In Section \ref{sec:transformedpayoff}, we will illustrate how we incorporate the changes in individual payoff to incorporate these mechanisms within our model of multilevel selection.

\subsubsection{Transformed Payoff Matrices and Multilevel Dynamics} 
\label{sec:transformedpayoff}

Here we present an approach we can use to incoporate the mechanisms of like-with-like assortment, other-regarding preferences, and reciprocity into our model of multilevel selection. For each of these mechanisms, we can describe the impact of these mechanisms by a one-parameter family of payoff functions $\pi_C^{\structure}(x)$ for $\pi_D^{\structure}(x)$, where the parameter $\structure$ will be depend on our model of the mechanism. We will associate each mechanism with a modfied family of payoff matrices that depend on our structure parameter $\structure$,

\begin{equation} \label{eq:transformedpayoffmatrix}
\begin{blockarray}{ccc}
& C & D \\
\begin{block}{c(cc)}
C & R_{\structure} & S_{\structure} \\
D & T_{\structure} & P_{\structure} \\
\end{block}
\end{blockarray}
\end{equation}
and we can use the transformed payoff matrix to calculate the effective payoffs $\pi_C^{\structure}(x) = R_{\structure}  x + S_{\structure} (1-x)$ and  $\pi_D^{\structure}(x) = T_{\structure}  x + P_{\structure} (1-x)$ for each mechanism \cite{nowak2006five,taylor2007transforming,kaznatcheev2018effective}. By further calculating the average payoff for group members $G_{\structure}(x) = x \pi_C^{\structure}(x) + (1-x) \pi_D^{\structure}(x)$, we can then incoporate each within-group mechanism into the dynamics of multilevel selection by considering solutions to the following PDE
\begin{equation} \label{eq:PDEmechanism}
\dsdel{f(t,x)}{t} = -\dsdel{}{x}\left[ x(1-x) \left( \pi_C^{\structure}(x) - \pi_D^{\structure}(x) \right) f(t,x) \right] + \lambda f(t,x) \left[G_{\structure}(x) - \int_0^1 G_{\structure}(y) f(t,y) dy \right].
\end{equation}
To explore the impact of each mechanism on the dynamics of multilevel selection, we will use both the modified payoff functions $\pi_C^{\structure}(x)$, $\pi_D^{\structure}(x)$, $G_{\structure}(x)$ and the modified payoff parameters $\alpha_{\structure} = R_{\structure} - S_{\structure} - T_{\structure} + P_{\structure}$, $\beta = S_{\structure} - P_{\structure}$, and $\gamma_{\structure} = S_{\structure} + T_{\structure} - 2 P_{\structure}$ to characterize the long-time support of cooperation. In particular, we will calculate the threshold selection strength $\lambda^*_{\structure}$, the average payoff at steady state $\langle G_{\structure}(\cdot) \rangle_{f^{\lambda}_{\theta}}$, and the modal composition of cooperation at steady state $\hat{x}_{\lambda}^{\structure}$, based on the formulas from Equations \eqref{eq:PDthresh}, \eqref{eq:steadystatefitness}, and \eqref{eq:PDmodal}, respectively.

\section{Within-Group Assortment} \label{sec:assortment}

In this section, we will consider the effect of within-group assortment on the dynamics of our PDE model of multilevel selection. We consider game-theoretic interactions within groups which follow an $r$-assortment process as introduced by Grafen and studied in various deterministic and stochastic settings \cite{grafen1979hawk,van2017hamilton,allen2015games,cooney2016assortment,coder2018effects,iyer2020evolution}. In particular, we assume a form of like-with-like assortment in which, with probability $r$, individuals play the game with an individual with the same strategy, while, with probability $1-r$, individuals play the game with a randomly chosen member of their group. In an infinitely large group with a fraction $x$ of cooperators, the expected payoff of a cooperator $\pi_C^r(x)$ and of a defector $\pi_D^r(x)$ under this assortment process is
\begin{subequations} \label{eq:rpayoffs} \begin{align}
\pi^r_C(x) &= r R + (1-r) \pi_C(x) =  r R + (1-r) (xR + (1-x)S) \\
\pi^r_D(x) &= r P + (1-r) \pi_D(x) =  rP + (1-r)(xT + (1-x) P)
\end{align}
\end{subequations}
We can also understand the role of the assortment process by describing the expected payoff with the following transformed payoff matrix for well-mixed interactions \cite{van2011replicator, nowak2006five}
\begin{equation} \label{eq:rpayoffmatrix}
\begin{blockarray}{ccc}
& C & D \\
\begin{block}{c(cc)}
C &  R  &  (1-r) S + r R  \\
D &  (1-r) T + r P & P  \\
\end{block}
\end{blockarray}.
\end{equation}
Using the shorthand $\alpha = R - S - T + P$, $\beta = S-P$, and $\gamma = S +T - 2P$, we see from Equation \eqref{eq:rpayoffs} that the difference in effective payoffs between cooperators and defectors is given by
\begin{equation} \label{eq:rwithingroup} %
 \pi^r_C(x)  - \pi^r_D(x) = %
 r\left( \gamma + \alpha \right) + \left(1 - r \right) \left(\beta + \alpha x \right) . \end{equation}
Using Equations \eqref{eq:withincharacteristics} and \eqref{eq:rwithingroup}, we see that the within-group dynamics have equilibria at $0$, $1$, and a third point $x_{eq}^r$ satisfying $\pi^r_C(x_{eq}^r) = \pi^r_D(x_{eq}^r)$, which is given by
\begin{equation} \label{eq:xeqr} x^{r}_{eq} = -\frac{\beta}{\alpha} + \left( \frac{r}{1-r} \right) \left(\frac{\gamma + \alpha}{-\alpha} \right).
 \end{equation}
 
The all-defector equilibrium is unstable when $\pi^r_C(0) > \pi^r_D(0)$, which, for an underlying PD game, occurs for $r$ exceeding the threshold value $r^s_W$ given by %
 \begin{equation} \label{eq:rsW} r^s_W = \frac{-\beta}{-\beta + \gamma + \alpha} \in (0,1) \end{equation}
Similarly, we define $r^a_W$ as the minimum level of $r$ above which $\pi^r_C(1) > \pi^r_D(1)$, which is given by
 \begin{equation} \label{eq:raW} r^a_W = \frac{- \left(\alpha + \beta\right)}{\gamma - \beta},  %
 \end{equation}
 and therefore the all-cooperator equilibrium is stable under the within-group dynamics when $r > r^a_W$. From Equations \eqref{eq:rsW} and \eqref{eq:raW}, we see that $r^s_W < r^a_W$ when $\alpha < 0$ and $r^s_W > r^a_W$ when $\alpha > 0$. This means that, as $r$ increases, the within-group dynamics will feature a stable intermediate equilibrum $x^r_{\mathrm{eq}}$ for $r^s_W < r < r^a_W$ (resembling the within-group Hawk-Dove dynamics) when $\alpha < 0$ and will feature bistability of the all-cooperator and all-defector equilibria (resembling the within-group Stag-Hunt dynamics) when $r^a_W < r < r^s_W$ when $\alpha > 0$.

Using Equation \eqref{eq:rpayoffs}, we see that the average payoff for a group with interactions following this $r$-assortment process is given by
\begin{align} G_r(x) &= x \pi^r_C(x) + (1-x) \pi^r_D(x) \nonumber \\ &= r\left[P + (R-P)x \right] + (1-r) \left[P + (S+T-2P)x + (R-S-T+P)x^2 \right] \\ &= r \left[P + \left( \gamma + \alpha \right) x \right] + \left(1 -r\right) \left[P + \gamma x + \alpha x^2 \right]. \label{eq:rbetweengroup} \end{align}
We see that the group payoff function interpolates between the group payoff function for well-mixed interations $G_0(x) = P +  \gamma x + \alpha x^2 = G(x)$ when $r = 0$ and an affine function of cooperator composition $G_1(x) = P + \left(\gamma + \alpha \right) x$ favoring as much cooperation as possible when $r = 1$. 

Noting that $G'_r(x) = r \left( \gamma + \alpha \right) + \left(1 - r \right) \left( \gamma + 2 \alpha x \right)$, we see that for PD games with average payoff $G(x)$ most favoring a mix of cooperators and defectors, average group payoff is maximized by the fraction of cooperators $x_r^*$ given by
 \begin{equation} \label{eq:xrstar}
   x_r^* = \left\{
     \begin{array}{cr}
       \ds\frac{\gamma}{-2\alpha} + \left( \frac{r}{1-r}\right) \left(\frac{\gamma + \alpha}{-2 \alpha} \right) & :r < r_B \vspace{1mm}\\
       1 & : r \geq r_B
     \end{array}
   \right.,
\end{equation} 
where the critical assortment parameter above which between-group competition most favors full-cooperator groups is \begin{equation} \label{eq:rB} r_B = \frac{\gamma + 2 \alpha}{\alpha}. \end{equation} %

From our characterization of the within-group dynamics in Equation \eqref{eq:rwithingroup} and group payoff function \eqref{eq:rbetweengroup}, we can now look to characterize how assortative interactions with assortment probability $r$ impacts the dynamics of multilevel selection. In Figure \ref{fig:rbifurcation}, we present two bifurcation diagrams illustrating the within-group equilibria and the collective optimum for cases of PD games in which group-level payoff is maximized by an intermediate level of cooperation under well-mixed interactions. From Equations \eqref{eq:rsW}, \eqref{eq:raW}, and \eqref{eq:rB}, we see that, when $\alpha < 0$, we do not have a definitive ordering on $r^s_W$ and $r_B$, so, in Figure \eqref{eq:rwithingroup}, we depict the two possible rankings of these threshold quantities: $r_B < r^s_W <  r^a_W$ (left) and $r^s_W < r_B <  r^a_W$ (right). In the former case, the  average group-level payoff $G_r(x)$ transitions to most favoring full-cooperation at an assortment probability for which the individual-level dynamics feature global stability of the all-defector equilibrium, while, in the latter case, we the within-group dynamics begin to feature a stable within-group equilibrium at an assortment probability for which group payoff is maximized by an intermediate mix of cooperators and defectors.

 In Table \ref{tab:rtable}, we characterize the long-time behavior of the multilevel dynamics for the different possible cases of the assortment probability $r$ and the payoff parameters $\gamma$ and $\alpha$ for the PD game. In addition to the case of $\alpha < 0$ illustrated in Figure \ref{fig:rbifurcation}, we also present the long-time behavior when $\alpha > 0$ and the within-group dynamics feature bistability of the all-cooperator and all-defector equilibria when $r_W^a < r < r_W^s$. The characterization of the long-time behavior for the different ranges of $\gamma$ and $\alpha$ is based on the characterizing the effective social dilemma given by the transformed payoff matrix of Equation \eqref{eq:rpayoffmatrix}, and then applying the result of Theorem \ref{thm:longtimesummary} that holds for the transformed game. 
 
 \begin{figure}[!ht]
 	\centering
 	\hspace{-5mm}\includegraphics[width=0.48\textwidth]{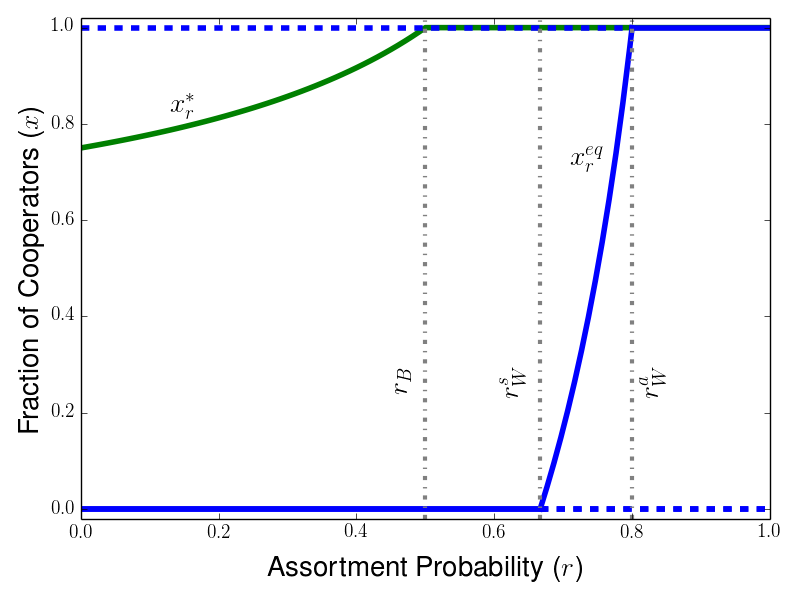}
 	\hspace{-5mm} \includegraphics[width=0.48\textwidth]{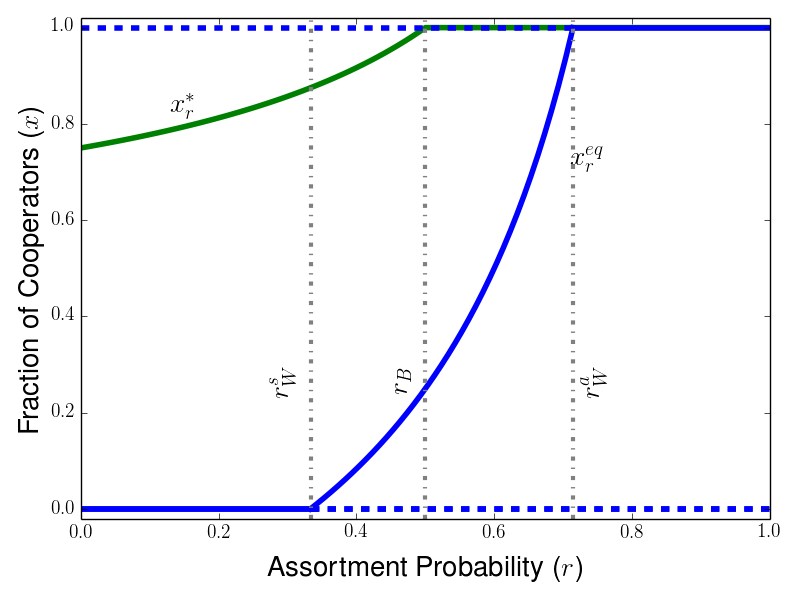}
	\caption{Bifurcation diagram for within-group replicator dynamics and group type with maximal average payoff our in model for other-regarding preferences. Game-theoretic parameters are given by $\gamma = 1.5$,  $\alpha = -1$, and either $\beta = -1$ (left) or $\beta = -1/4$ (right), displaying cases in which either $r_B < r_W^s$ (left) or $r_W^s < r_B$ (right). The green lines refer to the group composition which maximizes average payoff of group members $x^*_F$. Solid blue lines refer to stable equilibria of the within-group dynamics, while dashed blue lines describe unstable equilibria. The gray dot-dashed lines refer to the levels of the assortment probability above which the all-defector equilibrium becomes locally unstable $r_W^s$, above which the all-cooperator equilibrium becomes locally stable $r_W^a$, and above which the average payoff of group members is maximized by the all-cooperator group $r_B$.}
	\label{fig:rbifurcation}
 \end{figure}

\renewcommand{\arraystretch}{2}
\begin{table}[!ht]
\caption{Long-time behavior of within-group and multilevel dynamics for Prisoners' Dilemma games for various values of assortment probability $r$. The threshold $\lambda^*_r$ denotes the threshold strength of between-group competition $\lambda$ required to sustain cooperation at steady state when the multilevel dynamics with assortment are in the PD regime (corresponding to Equation \eqref{eq:PDthresh} for $\lambda^*_{PD}$). The threshold $\lambda^{**}_{r}$ corresponds to the analogous threshold when the effective payoffs of Equation \eqref{eq:rpayoffmatrix} correspond to a PD game, and it can be calculated using Equation \eqref{eq:HDthresh} for $\lambda^*_{HD}$.   }
\begin{center}
\begin{tabular}{|c|c|c|c|}
\hline
\makecell{Payoff \\ Parameters} & \makecell{Assortment \\  Probability ($r$)}  & Within-Group & Steady State for Multilevel Dynamics  \\
    \hline
  \multirow{3}{*}{\makecell{$\alpha < 0$}} & 
   $r < r_W^s$  & $0$ stable   & $\begin{aligned} \lambda \leq \lambda^*_r &: \textnormal{delta supported at } x=0  \\ \lambda > \lambda^*_r &: \textnormal{density supported on }  [0,1] \end{aligned}$   \\
  \cline{2-4}
   & $r_W^s < r < r_W^a $ &  $x^r_{eq}$ stable & $\begin{aligned} \lambda \leq \lambda^{**}_r &: \textnormal{delta concentrated at } x=x^r_{eq}  \\ \lambda > \lambda^*_r &: \textnormal{density supported on }  [x^r_{eq},1] \end{aligned}$  \\
   \cline{2-4}
 & $ r > r_W^a$ &  $1$ stable &  Delta concentrated at $x=1$ \\
   \hline
   \multirow{3}{*}{$\alpha > 0$} & 
   $r < r_W^s$   & $0$ stable  & $\begin{aligned} \lambda \leq \lambda^*_r &: \textnormal{delta supported at } x=0  \\ \lambda > \lambda^*_r &: \textnormal{density supported on }  [0,1] \end{aligned}$  \\
  \cline{2-4}
   & $r_W^a < r < r_W^{s}$ &  $0$, $1$ bistable & Delta concentrated at $x = 1$  \\
   \cline{2-4}
 & $ r > r_W^{s}$ & $1$ stable &  Delta concentrated at $x = 1$   \\
   \hline 
    
\end{tabular}
\end{center}
\label{tab:rtable}
\end{table}
\renewcommand{\arraystretch}{1}

Next, we can explore how the presence of assortment impacts the support for cooperation and the collective payoff achieved under the multilevel dynamics. When the assortment probability is sufficiently weak (when $r < r^s_W$ for $\alpha < 0$ and when $r < r^a_W$ for $\alpha > 0$), we can use the results for the multilevel PD dynamics to characterize the long-time outcome in terms of a tug-of-war between the collective incentive to cooperate $G_r(1) - G_r(0)$ and the individual incentive to defect in a many-cooperator group $\pi^r_D(1) - \pi^r_C(1)$. Using Equation \eqref{eq:rwithingroup}, we see that
\begin{equation} \label{eq:pidiffr}
\pi^r_D(1) - \pi^r_C(1) = - r (\gamma - \beta)  - (\beta + \alpha) = -r (T - P) + (T - R),
\end{equation}
so the individual incentive to defect is a decreasing function of $r$. From Equation \eqref{eq:rbetweengroup}, we see that group payoff function satisfies
\begin{equation} \label{eq:Gdiffr}
G_r(1) - G_r(0) = \gamma + \alpha \: \: \mathrm{and} \: \: G_r(1) = P + \gamma + \alpha = R = G(1),
\end{equation}
so the collective incentive to achieve full-cooperation over full-defection is unchanged by introducing assortative interactions. 

 Plugging these formulas for $\pi^r_D(1) - \pi^r_C(1)$ and $G_r(1) - G_r(0)$ into Equation \ref{eq:steadystatefitness}, we can find that the average payoff at steady state in the PD regime is \[ \langle G^r(\cdot) \rangle_{f^{\lambda}_{\theta}} = P + \gamma + \alpha + \frac{\left(\alpha + \beta + r \left( \gamma - \beta \right)  \right) \theta}{\lambda}, \]
so we can note that $\gamma - \beta = T - P > 0$ todeduce that the average payoff at steady state is increasing as assortment increases. Using Equation \ref{eq:PDthresh}, we can similarly find the threshold selection strength needed to achieve cooperation is given by
\[ \lambda^*_r = \frac{\left[-\left( \beta + \alpha  \right)+ \left( \beta - \gamma \right) r \right] \theta}{\gamma + \alpha},  \]
so the threshold to achieve cooperation decreases with increasing assortment probability $r$. As a result, we see that increasing the assortment probability both faciliates easier achievement of cooperation via multilevel selection as well as a greater long-time collective average payoff under the multilevel dynamics. However, noting from Equation \eqref{eq:Gdiffr} that $\lim_{\lambda \to \infty} \langle G^r(\cdot) \rangle_{f(x)}  = P + \gamma + \alpha = G(1) = R$, we find that assortment has no effect on the collective outcome achieved in the limit of infinite between-group selection strength. While adding assortment to the game-theoretic interactions does help for finite relative selection strength $\lambda$, we still see that best possible outcome of multilevel dynamics is still limited by the payoff of the all-cooperator group. Therefore adding assortment to the multilevel dynamics cannot erase  the shadow of lower-level selection, and an optimal intermediate mix of cooperation and defection still cannot be achieved under the multilevel dynamics with within-group assortment.

We can also examine how the assortment probability $r$ impacts the  steady state densities achieved via multilevel selection. In Figure \ref{fig:rdensities}, we display density steady states for the multilevel dynamics for various assortment probabilities $r$ and a fixed initial condition and relative selection strength $\lambda$. %
We see the densities support increasing levels of steady-state cooperation as $r$ increases, and that for the largest value of $r$ depicted, the within-group dynamics now favor a stable mix of cooperation depicted with a vertical dashed line with the same color as the corresponding density. %

 \begin{figure}[h]
 	\centering
 	\includegraphics[width=0.65\textwidth
 	]{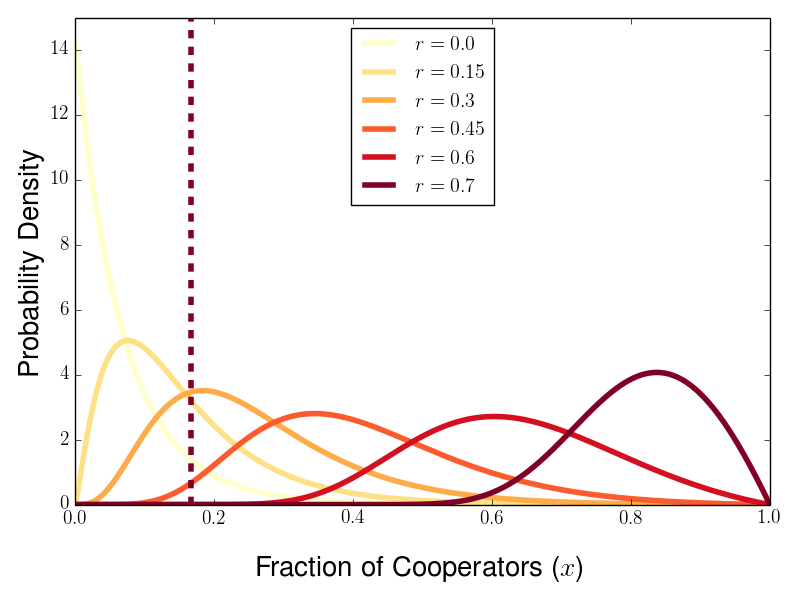}
	\caption{Steady state densities for $\lambda = 10$, $\gamma = 1.5$, $\alpha = \beta = -1$, $\theta = 2$ and various values of the assortment parameter $r$.  The vertical dotted line corresponds to the equilibrium for within-group dynamics for the value of $r = 0.7$, and is displayed in the same color as the corresponding steady state density for that assortment probability. }
	\label{fig:rdensities}
 \end{figure}

We can also consider the composition of group with greatest abundance at steady state, using Equation \eqref{eq:PDmodal} for the case of $r < r_W^a$ and the fact that the population concentrates upon full-cooperation for $r >  r_{W}^a$. In Figure \ref{fig:rpeakboth}, we show a comparison between the group type with maximal average payoff $x^*_r$ (given by Equation \eqref{eq:xrstar}) and the group type that is most abundant at steady state, plotted as a function of the assortment probability $r$. In the case of a fixed, finite $\lambda$ (Figure \ref{fig:rpeakboth},left), we see that the modal group type reaches full-cooperation as $r = r_W^a$ and the within-group dynamics themselves favor full-cooperation. In the case of infinite $\lambda$ (Figure \ref{fig:rpeakboth}, right), we see that the modal group type at steady state features fewer cooperators than in the optimal group $x^*_r$ when $r < r_B$, and then the optimal group and  modal group as $\lambda \to \infty$ coincide at full-cooperation when $r > r_B$. As a result, we see that incorporating additional within-group assortment helps to increase steady state cooperation for fixed $\lambda$ and to faciliate the achievement of all-cooperative outcomes in the limit of strong between-group competition.    %

    \begin{figure}[H]
 	\centering
 	\includegraphics[width=0.485\textwidth
 	]{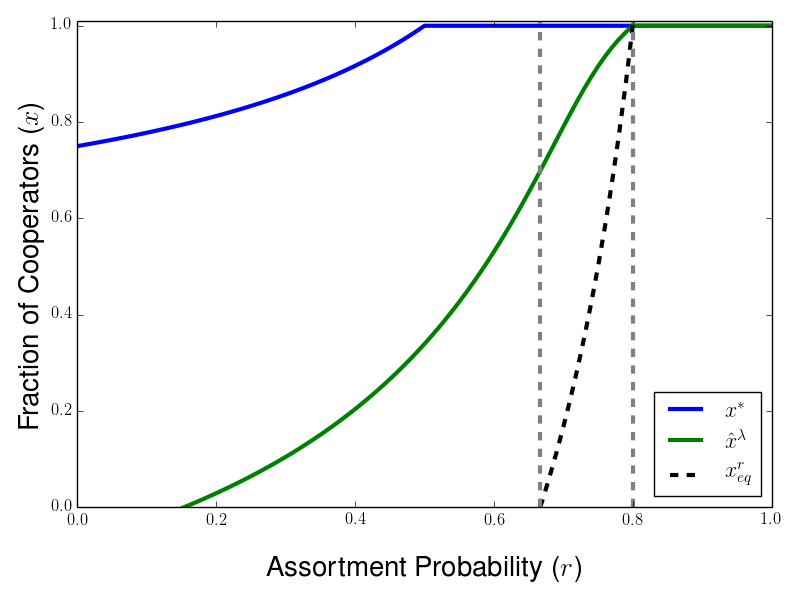}
 	\includegraphics[width=0.485\textwidth
 	]{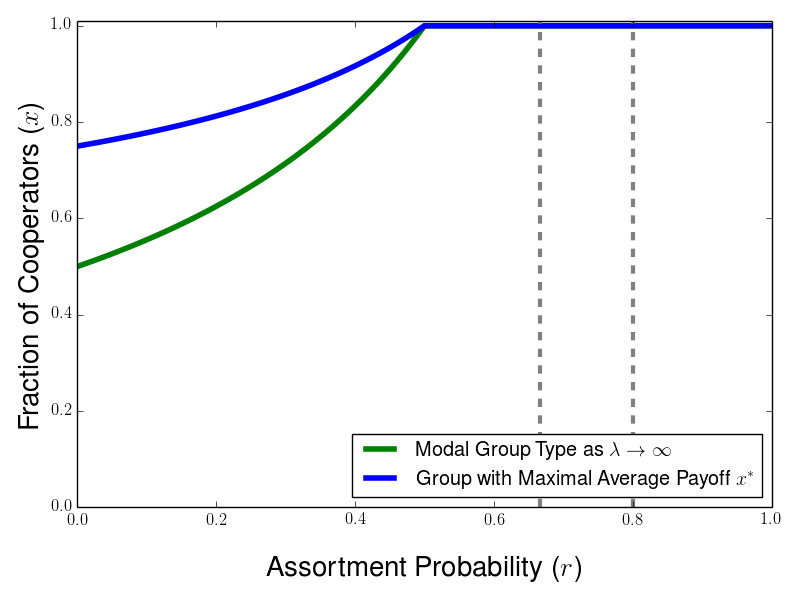}
	\caption{Group composition with maximal payoff $x^*$ (blue) and group type that is most abundant at steady state for relative strength of between-group selection $\lambda = 8$ (left) and in the limit as $\lambda \to \infty$ (right), plotted for various values of the assortment parameter $r$. Parameters for game are $\gamma = \frac{3}{2}$, $\alpha = \beta = -1$ for both panels, and the initial condition is chosen to satisfy $\theta = 2$ for the left panel. In both panels, left vertical dashed line corresponds to $r_W^s$ and right vertical dashed line corresponds to $r_W^a$. }
	\label{fig:rpeakboth}
 \end{figure}

 \section{Other-Regarding Preference} \label{sec:kinother}
 
 In this section, we will consider a model of multilevel selection in which individual-level replication rates depend not only on the payoffs received by a focal individual, but on a weighted average of the payoff of the focal player and their opponent. Weighing both the payoff of an individual and their opponent serves as a model of other-regarding preferences, as individuals take into account the effect of their cooperation or defection on others when evaluating their payoff for a given interaction. This mechanism of other-regarding preference is often used as a representation of the effect of genetic relatedness on cooperation in evolutionary games \cite{smith1982evolution,taylor2007transforming,szabo2012selfishness,pena2015evolutionary}, capturing the idea that individuals care not only about their own payoff but also the payoff of relatives.
 
 For our model of other-regarding preferences, we will assume that players interact with others according to the payoff matrix of Equation \eqref{eq:payoffmatrix}. From each interaction, the effective payoff obtained by a focal individual will be computed by placing weight $\frac{1}{1+F}$ on the individual's payoff and weight $\frac{F}{1+F}$ on the payoff of their opponent. The parameter $F$ measures the relative emphasis on one's own payoff and on the payoff of an opponent, and is sometimes referred to as the degree of fraternity for an interaction \cite{szabo2012selfishness}. For $F = 0$, individual-level success is determined only the payoff of the focal individual, while, for $F = 1$, individual success places equal weight on the payoff of the individual and their opponent.
 
 Using this rule for weighing the payoffs of an individual and their opponent, we see from the payoff matrix of Equation \eqref{eq:payoffmatrix} that, under  the effective payoffs of cooperators and defectors in an $x$-cooperator group are given by
 \begin{subequations} \label{eq:piF}
\begin{alignat}{2}
    \pi_C^F(x) &= x \left[  \frac{R}{1+F} + \frac{F R}{1+F}  \right] + \left( 1 -x \right)  \left[ \frac{S}{1+F}  + \frac{FT}{1+F}  \right]   &= \: Rx + \left( \frac{S + FT}{1+F} \right) (1-x) \\
    \pi_D^F(x) &= x  \left[ \frac{T}{1+F}  + \frac{F S}{1+F}  \right] + \left( 1 -x \right)  \left[ \frac{P}{1+F}  + \frac{FP}{1+F}   \right]  &=  \:  \left(\frac{T + FS}{1+F} \right) x + P(1-x)
    \end{alignat}
\end{subequations}
 
These effective payoffs can also be represented in terms of the following transformed payoff matrix:

 \begin{equation} \label{eq:otherregardingpayoffmatrix}
\begin{blockarray}{ccc}
& C & D \\
\begin{block}{c(cc)}
C &  R &  \ds \frac{S + F T}{1+F} \vspace{1mm} \\ 
D & \ds\frac{T + F S}{1+F} &  P \\
\end{block}
\end{blockarray} \: \: .
\end{equation}

Using our shorthand $\alpha = R - S - T + P$, $\beta = S-P$, and $\gamma = S+T-2P$, we can from Equation \eqref{eq:piF} that the difference between the effective payoffs of cooperators and defectors in an $x$-cooperator group are given by
\begin{dmath}
    \pi_C^F(x) - \pi_D^F(x) =  x \left[ R - \frac{S + FT}{1+F} - \frac{T + FS}{1+F} + P \right] + \frac{S+FT}{1+F} - P = \beta + \left(\frac{F}{1+F}\right) \left( \gamma - 2 \beta \right) + \alpha x. 
\end{dmath}
The full-defection equilibrium becomes unstable when $\pi_C^F(0) > \pi_D^F(0)$, which occurs for
\begin{equation} \label{eq:Fws}
F > F_W^s := \frac{-\beta}{\gamma - \beta}.
\end{equation}
We see that $F_W^s < 1$ provided that $\gamma > 0$, while the all-defector equilibrium will remain stable for all $F$ for underlying PD games in which $\gamma < 0$. This stands in contrast to the $r$-assortment model, in which the all-defector equilibrium is always destabilized for sufficiently strong assortment probability.

The full-cooperator equilibrium becomes stable when $\pi_C^F(1) > \pi_D^F(1)$, which occurs for
\begin{equation} \label{eq:Fwa}
F > F_W^a := \frac{-\left(\beta + \alpha\right)}{\gamma + \alpha - \beta}. \end{equation}
From Equations \ref{eq:Fws} and \ref{eq:Fwa}, we see that $F_W^s < F_w^a$ for $\alpha < 0$, while $F_W^s > F_w^a$ for $\alpha > 0$. 

For a range of fraternity parameters $F$, there also exists an interior equilibrium $x_{eq}^F$ at which $\pi_C^F(x_{eq}^F) = \pi_D^F(x_{eq}^F)$, which is given by
\begin{equation} \label{eq:xeqF}
x_{eq}^F = \frac{\beta}{-\alpha} + \left(\frac{F}{1+F}\right) \left( \frac{\gamma - 2 \beta}{-\alpha} \right).
\end{equation}
This equilibrium is stable when it exists when $\alpha < 0$, while it unstable and seperates the basins-of-attraction for full-defection and full-cooperation when $\alpha > 0$. 

When $F= 1$, the payoff difference of Equation \eqref{eq:piF} takes the form
\begin{dmath}
    \pi_C^1(x) - \pi_D^1(x)  = \frac{\gamma}{2} + \alpha x, 
\end{dmath}
and so we see that, in this case, within-group dynamics can support global stability of full-cooperation (when $\gamma > 0$ and $\gamma + 2 \alpha > 0$), bistability of full-cooperation and full-defection (for PD games with $\gamma < 0$, whose payoff parameters must correspondingly satisfy $\alpha > 0$ and $\gamma + 2 \alpha > 0$), or stability of an interior equilibrium (when $\gamma > 0$ and $\gamma + 2 \alpha < 0$). This stands in contrast to the case of the $r$-assortment model, which only allows for the possibility of global stability of full-cooperation under individual-level selection when $r = 1$. 

In addition, we see from taking the limit as $F \to 1$ in Equation \eqref{eq:xeqF} that, in this case, the interior equilibrium takes the form
\begin{equation} \label{eq:xeqF1}
x_{eq}^F \bigg|_{F = 1} = \frac{\gamma}{-2 \alpha} = x^*,
\end{equation}
which is the intermediate level of cooperation $x^*$ that maximizes average payoff for a group playing a PD game when $\gamma + 2 \alpha < 0$. Consequently, when individuals care just as much about their own payoff as the payoff for their opponent, within-group selection can promote the level of cooperation that maximizes the average payoff of the population for the underlying PD game. 

Next, we consider the role of average group payoff in this model with other-regarding preferences. We compute that the average payoff of group members for an $x$-cooperator group is given by
\begin{dmath}
G_F(x) = x \pi_C^F(x) + (1-x) \pi_D^F(x) = x \left[ Rx + \left( \frac{S + FT}{1+F} \right) (1-x) \right] + (1-x) \left[\left(\frac{T + FS}{1+F} \right) x + P(1-x) \right] =\underbrace{ P + \left( S + T - 2P \right)x + (R-S-T+P) x^2}_{= G(x)}.
\end{dmath}
Therefore the collective average payoff is unchanged as $F$ ranges between $0$ and $1$. In particular, this means that if the baseline game has a group payoff function $G(x)$ with an intermediate optimum, then $G_F(x)$ also has collective payoff maximized by the same intermediate level of cooperation.

Because the group-reproduction function $G_F(x)$ does not depend on $F$, we further see in the case that $G_F(x)$ is maximized by an interior level of cooperation that $G(x_{eq}^F)$ will exceed $G(1)$ when $x_{eq}^F \in \left(\frac{\gamma+\alpha}{-\alpha}, 1 \right)$. In this case, $x_{eq}^F$ is globally stable under the within-group dynamics, while the condition $G_F(x_{eq}^F) > G_F(1)$ reflect multilevel dynamics consistent with the anti-coordination games (AC1,AC2) rather than a Hawk-Dove game. The condition $G(x_{eq}^F) > G(1)$  is satisfied when $\gamma + 2 \alpha < 0$ and $F$ exceeds the following threshold level
\begin{equation} \label{eq:FwAC}
F > F_W^{AC} := \frac{\gamma + \alpha - \beta}{-(\beta+\alpha)} = \frac{1}{F_W^a}.
\end{equation}
Notably, only one of the thresholds $F_W^{AC}$ or $F_W^{a}$ can be achieved by a feasible weight $F \leq 1$ placed on one's opponent's payoff. We can deduce from Equations \ref{eq:Fwa} and \ref{eq:FwAC} that $F_W^{a} < 1$ when $G_F(x) = G(x)$ is maximized by full-cooperation ($\gamma + 2 \alpha > 0$), while $F_W^{AC} < 1$ when $G_F(x)$ has an interior optimum ($\gamma + 2 \alpha < 0$).

Now that we have characterized the within-group and between-group replication rates for our model with other-regarding preferences, we illustrate several possible bifurcation diagrams for our model in Figure \ref{fig:Fbifurcation} as we vary the fraternity parameter $F$. In particular, we illustrate a case in which the full-cooperator equilibrium $x=1$ maximizes average group payoff $G_F(x)$ (Figure \ref{fig:Fbifurcation}, left), and we see that the within-group dynamics eventually favor full-cooperation for $F$ sufficiently close to 1. We also illustrate a case in in which average payoff is maximized by an intermediate fraction of cooperators (\ref{fig:Fbifurcation}, right), and we see that, as $F \to 1$, the equilibrium $x_{eq}^F$ converges to the intermediate social optimum. %
In Table \ref{tab:Ftable}, we characterize the long-time behavior of the individual-level and multilevel selection for the different possible combinations of the payoff parameters $\gamma$ and $\alpha$ and the fraternity parameter $F$. This highlights the broader class of dynamical behaviors that are possible for the model with other-regarding preferences, as compared to the $r$-assortment model and the models of reciprocity that we will present in Section \ref{sec:reciprocity}.

\begin{figure}[ht]
 	\centering
 	\hspace{-5mm}\includegraphics[width=0.48\textwidth]{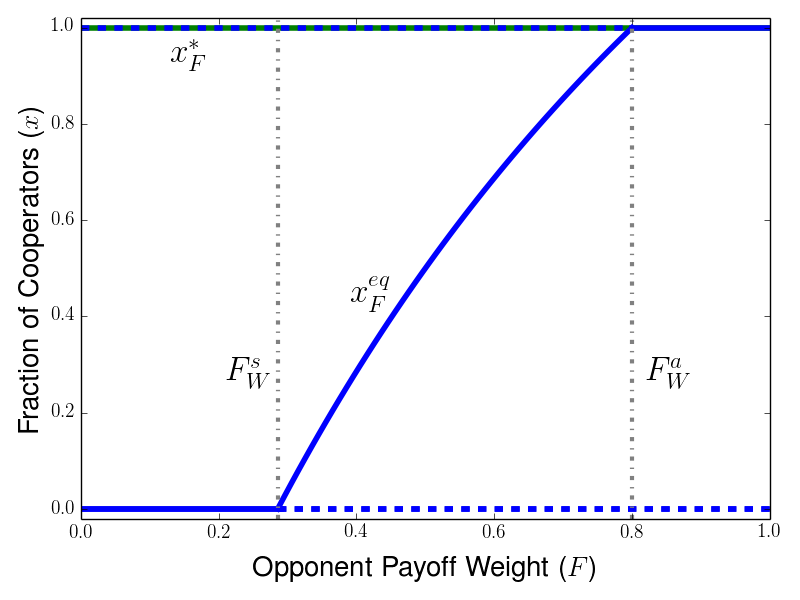}
 	\hspace{-5mm} \includegraphics[width=0.48\textwidth]{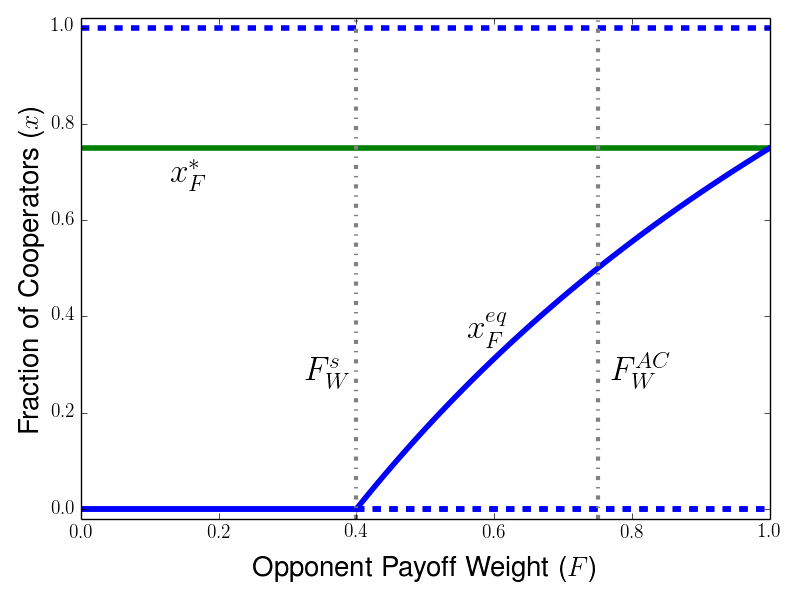}
	\caption{Bifurcation diagram for within-group replicator dynamics and group type with maximal average payoff in assortment model. The green lines refer to the group composition which maximizes average payoff of group members $x^*_F$. Solid blue lines refer to stable equilibria of the within-group dynamics, while dashed blue lines describe unstable equilibria. The gray dot-dashed lines refer to the fraternity parameter above which the all-defector equilibrium becomes locally unstable $F_W^s$, above which the all-cooperator equilibrium becomes locally stable $F_W^a$, and above which the effective payoff reflects an anti-coordination game $F_W^{AC}$. }
	\label{fig:Fbifurcation}
 \end{figure}

\renewcommand{\arraystretch}{2}
\begin{table}[ht!]
\caption{Long-time behavior of within-group and multilevel dynamics for Prisoners' Dilemma games with other-regarding preferences. The threshold $\lambda^*_F$ denotes the threshold strength of between-group competition $\lambda$ required to sustain cooperation at steady state when the multilevel dynamics with other-regarding preferences are in the PD regime (corresponding to Equation \eqref{eq:PDthresh} for $\lambda^*_{PD}$). The threshold $\lambda^{**}_{F}$ corresponds to the analogous threshold when the effective payoffs of Equation \eqref{eq:rpayoffmatrix} correspond to an HD game, and it can be calculated using Equation \eqref{eq:HDthresh} for $\lambda^*_{HD}$. }
\begin{center}
\begin{tabular}{|c|c|c|c|} 
\hline
\makecell{Payoff \\ Parameters} & \makecell{Other-Regarding \\  Preference ($F$)}  & Within-Group & Steady State for Multilevel Dynamics  \\
    \hline
  \multirow{3}{*}{\makecell{$\gamma > 0$, $\alpha < 0$, \\$\gamma + 2 \alpha < 0$}} & 
   $F < F_W^s$   & $0$ stable & $\begin{aligned} \lambda \leq \lambda^*_F &: \textnormal{delta supported at } x=0  \\ \lambda > \lambda^*_F &: \textnormal{density supported on }  [0,1] \end{aligned}$ \\
  \cline{2-4}
   & $F_W^s < F < F_W^{AC}$  & $x_{eq}^F$ stable &  $\begin{aligned} \lambda \leq \lambda^{**}_F &: \textnormal{delta concentrated at } x=x_{eq}^F  \\ \lambda > \lambda^*_F &: \textnormal{density supported on }  [x_{eq}^F,1] \end{aligned}$ \\
   \cline{2-4}
 & $ F > F_W^{AC}$ & $x_{eq}^F$ stable & Delta concentrated at $x = x_{eq}^F$   \\
   \hline
  \multirow{3}{*}{\makecell{$\gamma > 0$, $\alpha < 0$, \\ $\gamma + 2 \alpha > 0$}} & 
   $F < F_W^s$  & $0$ stable   & $\begin{aligned} \lambda \leq \lambda^*_F &: \textnormal{delta supported at } x=0  \\ \lambda > \lambda^*_F &: \textnormal{density supported on }  [0,1] \end{aligned}$   \\
  \cline{2-4}
   & $F_W^s < F < F_W^{a}$ &  $x_{eq}^F$ stable & $\begin{aligned} \lambda \leq \lambda^{**}_F &: \textnormal{delta concentrated at } x=x_{eq}^F  \\ \lambda > \lambda^*_F &: \textnormal{density supported on }  [x_{eq}^F,1] \end{aligned}$  \\
   \cline{2-4}
 & $ F > F_W^{a}$ &  $1$ stable &  Delta concentrated at $x=1$ \\
   \hline
   \multirow{3}{*}{$\gamma, \alpha > 0$} & 
   $F < F_W^s$   & $0$ stable  & $\begin{aligned} \lambda \leq \lambda^*_F &: \textnormal{delta supported at } x=0  \\ \lambda > \lambda^*_F &: \textnormal{density supported on }  [0,1] \end{aligned}$  \\
  \cline{2-4}
   & $F_W^s < F < F_W^{a}$ &  $0$, $1$ bistable & Delta concentrated at $x = 1$  \\
   \cline{2-4}
 & $ F > F_W^{a}$ & $1$ stable &  Delta concentrated at $x = 1$   \\
   \hline 
      \multirow{2}{*}{$\gamma <  0$, $\alpha > 0$} & $F < F_W^a$ &   $0$ stable   & $\begin{aligned} \lambda \leq \lambda^*_F &: \textnormal{delta supported at } x=0  \\ \lambda > \lambda^*_F &: \textnormal{density supported on }  [0,1] \end{aligned}$   \\
  \cline{2-4} 
    & $F > F_W^a$ &  $0$, $1$ bistable & Delta concentrated at $x = 1$  \\
  \hline
    
\end{tabular}
\end{center}
\label{tab:Ftable}
\end{table}
\renewcommand{\arraystretch}{1}

Under this model for other-regarding preferences, the individual-level advantage of defecting in a many-cooperator group is given by
\begin{equation}
    \pi_D^F(1) - \pi_C^F(1) = -(\beta +  \alpha) - \left( \frac{F}{1+F} \right) \left( \gamma - 2 \beta\right),
\end{equation}
which is a decreasing function of the weight $F$ placed on other-regarding payoff (as $\gamma - 2 \beta = T - S > 0$ for any PD game). The collective incentive to cooperate is given by $G_F(1) - G_F(0) = G(1) - G(0) = \gamma + \alpha$, which is constant in $F$. When $F < \min(F_W^s,F_W^a)$ and the effective game under other-regarding preference corresponds to a PD game, we can use these formulas and Equation \eqref{eq:PDthresh} to see that cooperation can be achieved via multilevel selection when the relative strength of between-group selection satisfies
\begin{equation} \label{eq:lambdastarF}
\lambda > \lambda^*_F := \frac{- \left[ \beta + \alpha + \left( \frac{F}{1+F} \right) \left( \gamma - 2 \beta\right)\right] \theta}{\gamma + \alpha}.
\end{equation}
In addition, we can use Equation \eqref{eq:steadystatefitness} to see that, for this range of $F$, the average payoff at steady state is given by
\begin{equation}
\langle G_F(\cdot) \rangle_{f^{\lambda}_{\theta}} = \gamma + \alpha + \frac{ \left[ \beta + \alpha + \left( \frac{F}{1+F} \right) \left( \gamma - 2 \beta\right)\right] \theta}{\lambda \left(\gamma + \alpha\right)}.
\end{equation}
Therefore we see that, for $F < \min(F_W^s,F_W^a)$, $\lambda^*_F$ is decreasing in $F$ and $\langle G_F (\cdot) \rangle_{f^{\lambda}_{\theta}}$ is increasing in $F$, so increasing the weight individuals place on the payoff of their opponents makes it easier to achieve cooperation via multilevel selection and to achieve a higher collective payoff at steady state.

In Figure \ref{fig:Fdensities}, we display how changing the fraternity parameter $F$ impacts the steady state densities $f^{\lambda}_{\theta}(x)$ achieved as the long-time behavior for a given initial density, fixed relative selection strength $\lambda$, and a pair of PD games with $\gamma > 0$ and $\alpha < 0$. The expression for these densities are derived using the effective payoff matrix of Equation \eqref{eq:otherregardingpayoffmatrix}, as well as the formulas from Equation \eqref{eq:PDsteady} and \eqref{eq:HDsteady} for the cases in which $F < F_W^s$ and in which $F_W^s < F < \min(F_W^a, F_W^{AC})$, respectively. In Figure \ref{fig:Fdensities}(left), we illustrate a case in which full-cooperation is collectively optimal ($\gamma + 2 \alpha > 0$), showing that increasing levels of weight $F$ placed on other-regarding preference allows steady state outcomes approaching the full cooperation outcome as $F \to F_W^a$. In Figure \ref{fig:Fdensities}(right), we depict a case in which average group payoff is maximized by 75 percent cooperation, and see that the steady states appear to concentrate upon a concentration of 50 percent cooperation as $F \to F_W^{AC}$. For this choice of parameters, the steady state population is approaching the composition with the same collective payoff as the full-cooperator group, consistent with the fact the multilevel dynamics will concentrate upon a delta function $\delta(x - x_{eq}^F)$ when $F > F_W^{AC}$ and the effective payoff matrix resembles an anti-coordination game.

\begin{figure}[ht!]
 	\centering
 	\hspace{-5mm}\includegraphics[width=0.48\textwidth]{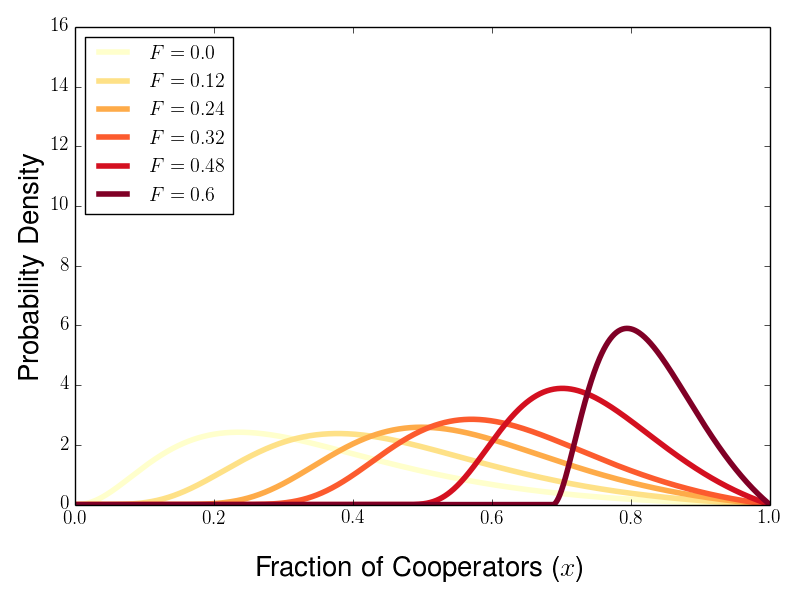}
 	\hspace{-5mm} \includegraphics[width=0.48\textwidth]{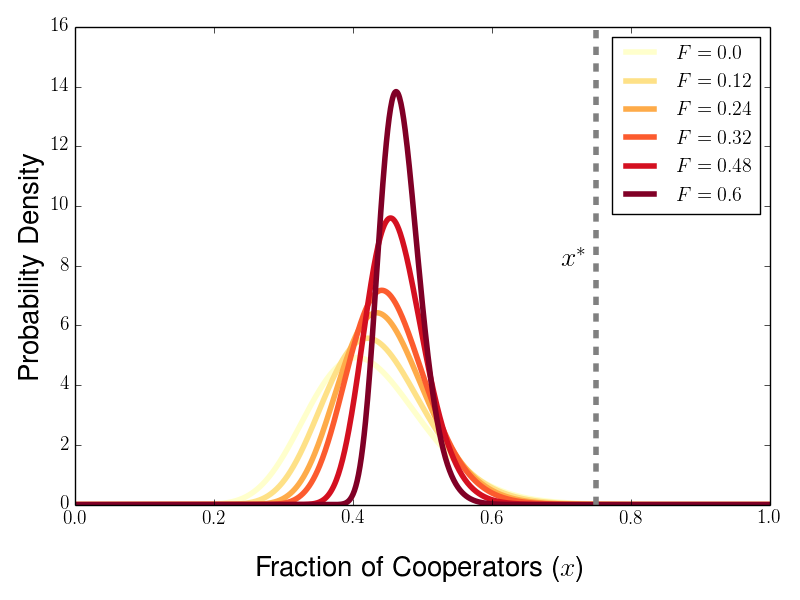}
	\caption{Steady-state densities for various weights of opponent payoff $F$ and fixed relative between-group selection strength $\lambda = 5$.  The parameters to generate these steady states are $\gamma = 2.5$ (left) and $\gamma = 1.5$ (right), with $\beta = \alpha = -1$ and $\theta = 2$ for both panels. The black dashed line in the right panel corresponds to the composition $x^*_F = \tfrac{\gamma}{-2\alpha}$ that optimizes group payoff.}
	\label{fig:Fdensities}
 \end{figure}
 
 In Figure \ref{fig:Fpeak}, we present the modal fraction of cooperators $\hat{x}^{\lambda}_F$ achieved at steady state across the range of possible fraternity parameters $F \in [0,1]$, given a fixed initial density, relative strength of between-group competition $\lambda$, and underlying PD game for which average group payoff is maximized by a composition with 75 percent cooperators. This modal outcome is calculated from Equation \eqref{eq:PDmodal} when the long-time dynamics result in a bounded steady state density, and coincide with the within-group equilibrium $x_{eq}^F$) for values of $F$ in which the long-time behavior concentrates upon the delta-function $\delta(x-x_{eq}^F)$. We see that the modal cooperation is increasing in the fraternity parameter $F$, and that this modal value approaches the socially-optimal level $x^*_F = \frac{\gamma}{-2\alpha} = x^*$ as $F \to 1$. Because the group payoff function $G_F(x) = G(x)$ is unchanged by incorporating other-regarding preferences, the corresponding concentration of the steady-state population upon the socially-optimal level of cooperation $x^*_F$ means that collective payoff population will eventually exceed $G(1)$ and achieve the optimal value $G(x^*) = G(\frac{\gamma}{-2\alpha})$. In this sense, we can see the other-regarding preference as a mechanism that helps to overcome the shadow of lower-level selection. 
 
 \begin{figure}[ht!]
 \centering
 \includegraphics[width = 0.65\textwidth]{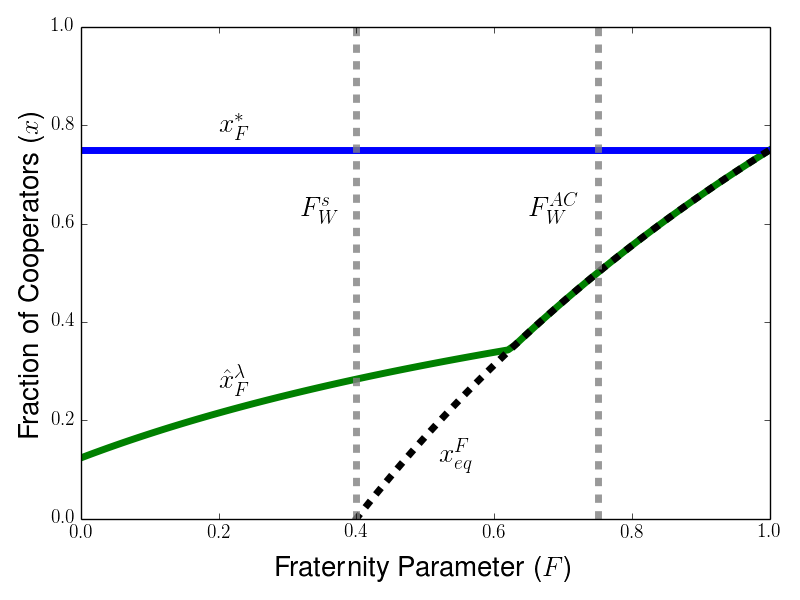}
 \caption{Comparsion of modal level of cooperation achieved at steady state with the composition of cooperators maximizing average payoff, plotted as a function of the fraternity parameter $F$. The solid blue line represents the maximal group payoff $x^*_F = \tfrac{\gamma}{-2\alpha}$, the solid green line represents the modal composition at steady state $\hat{x}^{\lambda}_F$, the dashed black line represents the interior within-group equilibrium $x_{eq}^F$ when it exists, and the vertical dashed gray lines describe the fraternity levels $F_W^s$ and $F_W^{AC}$ at which the effective payoffs transition to an HD game and to an AC game, respectively. }
  \label{fig:Fpeak}
 \end{figure}

Intuitively, we see that this ability for within-group competition to achieve a social optimum stems from the fact that for $F = 1$, the payoffs received by individuals in an interaction are equal to half of the payoff contributed to the average payoff of group members generated by this interaction. As a result, this equal weighting of payoffs perfectly aligns the individual-level and group-level interests derived from the effective payoffs from the game-theoretic interactions, and synchronizes the dynamics of selection at the two levels upon favoring the outcome $x^* = \frac{\gamma}{-2\alpha}$ when $F=1$. This stands in contrast to the mechanism of the $r$-assortment process, which serves to cluster cooperators with cooperators and defectors with defectors, faciliating increased cooperation but with no effort to promote interact the cooperator-defector interactions that can generate optimal group benefit when $S+T > 2R$ (and correspondingly $\gamma + 2 \alpha < 0$). %
 
 \section{Reciprocity Mechanisms} \label{sec:reciprocity}
 
 Now we turn to analyzing interactions with reciprocity mechanisms, which punish defectors for their reputation of bad behavior and can help to stabilize populations with an established social norm of cooperation. In Section \ref{sec:indirectreciprocity},  we consider a model of Nowak and Sigmund for social interactions with indirect reciprocity in which cooperators sometimes detect that their opponent is a defector and punish them for their defector status \cite{nowak2005indirect}. In Section \ref{sec:directreciprocity}, we consider a model of direct reciprocity via repeated games, exploring multilevel competition between individuals who always defect and those who conditionally cooperation following the tit-for-tat or grim trigger strategies.   %
 
 \subsection{Simplified Model for Indirect Reciprocity} \label{sec:indirectreciprocity}

In this section, we consider the simplifed model of indirect reciprocity introduced by Nowak and Sigmund \cite{nowak1998evolution}, which modifies our baseline model of game-theoretic interactions by allowing for the possibility for cooperative individuals to punish defection. In particular, we retain the assumption that defectors will always defect, and that cooperators will always cooperate in interactions with other cooperators. The main feature of this model is our assumption about the behavior of cooperators when paired with a defector: with probability $q$, they recognize the defector and punish them with defection, while, with probability $1-q$, they do not recognize the defector and choose to cooperate.  If the payoffs from these interactions follow the payoff matrix of Equation \eqref{eq:payoffmatrix}, then the interaction rules for this model of reciprocity result in the following effective payoffs for cooperators and defectors in an $x$-cooperator group
\begin{subequations} \label{eq:piqx}
\begin{alignat}{2}
\pi^q_C(x) &= xR + (1-x) \left(q P + (1-q) S \right) &&= \left( 1 - q\right) \pi_C(x) + q P + q(\gamma + \alpha)x \\
\pi^q_D(x) &= x \left(q P + (1-q) T \right) + (1-x) P && =  \left(1 - q \right) \pi_D(x) + q P.
\end{alignat}
\end{subequations}
For this process, it can also be helpful to view the detection probability $q$ and the expected payoffs in terms of a transformed payoff matrix  
\begin{equation} \label{eq:qpayoffmatrix}
\begin{blockarray}{ccc}
& C & D \\
\begin{block}{c(cc)}
C &  R  &  (1-q) S + q P   \\
D & (1-q) T + q P & P  \\
\end{block}
\end{blockarray}
\end{equation}
\cite{taylor2007transforming}. From Equation \eqref{eq:piqx}, we see that the difference in payoffs between a cooperator and defector in the same group is given by
\begin{equation} \label{eq:pidiffq}
\pi_C^q(x) - \pi_D^q(x) = \left(1-q\right) \beta + \left( \alpha + q \gamma \right) x.
\end{equation}
By plugging in $x=0$, we see from Equation \eqref{eq:pidiffq} that $\pi_C(0) - \pi_D(0) = \beta < 0$, so the all-defector equilibrium will be locally asymptotically stable for any $q < 1$. In addition, we see that, for any PD game, the all-cooperator equilibrium will become locally asymtotically stable when $\pi_C(1) > \pi_D(1)$, which occurs for defector detection probability satisfying %
\begin{equation} \label{eq:qwg} q > q_{WG} := \frac{- \left( \beta + \alpha \right)}{\gamma - \beta}  \end{equation}
This means that, for $q \in (q_{WG},1)$, the all-cooperator and all-defector equilibria will be bistable under the within-group dynamics, separated by the unstable intermediate equilibrium given by
\begin{equation} \label{eq:xeqq}
x_{eq}^q =  - \frac{(1-q)\beta}{\alpha + q \gamma}.
\end{equation}
Average payoff in a group with fraction $x$ cooperators is given by 
\begin{equation} 
\begin{aligned}
G_q(x) &= x \left[(1-q) \pi_C(x) + q P  + q(\gamma + \alpha) x  \right] + (1-x) \left[(1-q) \pi_D(x) + qP \right] \\ 
&= P  + (1-q) \gamma x +  (q \gamma + \alpha) x^2.
\end{aligned}
\end{equation}

For PD games with intermediate group payoff maxima $x^* = \frac{\gamma}{-2\alpha} < 1$, we see that there is a threshold level of detection probability $q_{BG}$ given by
\begin{equation} \label{eq:qbg} q_{BG} = \frac{-2 \alpha}{\gamma} - 1 \end{equation} 
such that the level of cooperation $x^*_{q}$ maximizing collective payoff $G_q(x)$ under indirect reciprocity has the piecewise characteriation 
\begin{equation} \label{eq:xstarq}
x^*_q = \left\{
     \begin{array}{cr}
      \ds \frac{(1-q) \gamma}{- 2\left(\alpha + q \gamma\right)} & : q < q_{BG} \vspace{1mm} \\
       1 & : q \geq q_{BG}
     \end{array}
   \right. .
\end{equation}

We note that
\[ q_{WG} - q_{BG} = \frac{\left( \gamma + \alpha \right) \left( \gamma - 2 \beta\right)}{\gamma \left( \gamma - \beta \right)} < 0\]
for all PD games, so full-cooperation will always be most favored by between-group competition for values of the punishment probability $q$ for which the all-cooperator equilibrium is locally stable under within-group competition. As a result, the effective payoff matrix of Equation \eqref{eq:qpayoffmatrix} will correspond to an SH game with $q > q_{WG}$, so, for such values of $q$, the population will concentrate upon a delta-function at full-cooperation in the presence of any between-group competition (i.e. when $\lambda > 0$). 

From the above properties of individual and collective payoff functions $\pi_C^q(x)$, $\pi_D^q(x)$, and $G_q(x)$, we are able to illustrate in Figure \ref{fig:qbifurcation} the generic bifurcation for the two-level q-process dynamics for an underlying PD game with intermediate group optimum. In particular, we see as $q$ increases that optimal group composition $x^*_q$ increases to $1$ as $q$ increases to $q_{BG}$, while within-group dynamics still favor full-defection in this regime and the multilevel dynamics still resemble the PD. Then, as $q$ increases past $q_{WG}$, the unstable equilibrium $x_{eq}^q$ appears, allowing for the bistability of full-cooperation and full-defection under the within-group dynamics. Finally, full-cooperation becomes globally stable within groups when $q = 1$ and cooperators punish defectors whenever they interact. In Table \ref{tab:qtable}, we further characterize the behavior of our model of indirect reciprocity under individual-level and multilevel selection, seeing that the effective payoff matrix of Equation \eqref{eq:qpayoffmatrix} corresponds to a PD game for $q < q_{WG}$ and an SH game for $q > q_{WG}$.   %
 \begin{figure}[ht!]
 	\centering
 	\hspace{-5mm}\includegraphics[width=0.6\textwidth
 	]{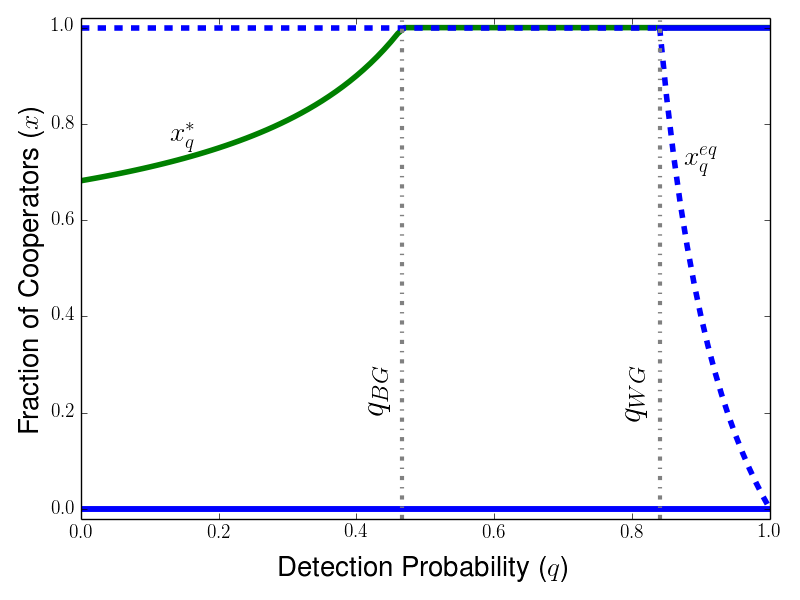}
	\caption[Bifurcation diagram for within-group replicator dynamics and group type with maximal average payoff for indirect reciprocity model.]{Bifurcation diagram for within-group replicator dynamics and group type with maximal average payoff for indirect reciprocity model. Green line describes group type $x^*_q$ with maximal average payoff of group members $G_q(x)$. Solid blue lines describe stable equilibria of within-group dynamics, while dashed blue lines correspond to unstable equilibria. The leftmost dot-dashed gray line corresponds to $q_{BG}$, the detection probability above which group payoff $G_q(x)$ is best off with full cooperation, and the rightmost gray line corresponds to $q_{WG}$, the detection probability above which full-cooperation is locally stable within groups. }
	\label{fig:qbifurcation}
 \end{figure}

 \renewcommand{\arraystretch}{2}
\begin{table}[ht!]
\caption{Long-time behavior of within-group and multilevel dynamics for Prisoners' Dilemma games with indirect reciprocity for various values of defector detection probability $q$. The threshold $\lambda^*_q$ denotes the threshold strength of between-group competition $\lambda$ required to sustain cooperation at steady state when the multilevel dynamics with indirect reciprocity are in the PD regime (corresponding to Equation \eqref{eq:PDthresh} for $\lambda^*_{PD}$).}
\begin{center}
\begin{tabular}{|c|c|c|}
\hline
 \makecell{Detection \\  Probability ($q$)}  & \makecell{Within-Group \\ Dynamics} & Steady State for Multilevel Dynamics  \\
    \hline
    $q < q_{WG}$ & $0$ stable & $\begin{aligned} \lambda \leq \lambda^*_q &: \textnormal{delta supported at } x=0  \\ \lambda > \lambda^*_q &: \textnormal{density supported on }  [0,1] \end{aligned}$ \\
\hline 
 $q_{WG} < q < 1$ & $0,1$ bistable & Delta concentrated upon $x=1$ \\
   \hline 
    
\end{tabular}
\end{center}
\label{tab:qtable}
\end{table}
\renewcommand{\arraystretch}{1}

Using these formulas, we see that $G(1) = \gamma + \alpha$ and that $\pi_D^q(1) - \pi_C^q(1) = - \left( \beta + \alpha\right) - q \left( \gamma - \beta \right)$. When $q < q_{WG}$ and the multilevel dynamics are in the PD regime, we then see from Equation \ref{eq:PDthresh} that the threshold level of relative selection strength needed to achieve cooperation is 
\begin{equation} \label{eq:lambdastartq} \lambda^*_q = \frac{\theta}{\gamma + \alpha} \left(  - \left( \beta + \alpha\right) - q \left( \gamma - \beta \right) \right),  \end{equation}
which is decreasing in $q$. Further, we see that \[ \lim_{q \to q_{WG}} \lambda^*_q =  \frac{\theta}{\gamma + \alpha} \left[- (\beta + \alpha) - \left(\frac{-(\beta + \alpha)}{\gamma - \beta} \right) \left( \gamma - \beta \right) \right]= 0, \] so the relative strength of between-group competition $\lambda^*_q$ needed to achieve any cooperation decreases to $0$ in the limit as $q \to q_{WG}$. This coincides with $q$ achieving the punishment probability at which within-group selection is sufficient to sustain the stability of cooperation on its own.

We can also use Equation \ref{eq:steadystatefitness} to see that the average payoff at the steady state density is \[ \langle G^q(\cdot) \rangle_{f^{\lambda}_{\theta}} = \gamma + \alpha + \frac{\left(\alpha + \beta + q \left( \gamma - \beta \right)  \right) \theta}{\lambda}, \]
so the average payoff at steady state is increasing as the probability of detecting defectors. Further, we see that as $q \to q_{WG} = -\frac{\alpha + \beta}{\gamma - \beta}$ that $ \langle G^q(\cdot) \rangle_{f^{\lambda}_{\theta}(x)} = \gamma + \alpha = G(1)$, in agreement with the observation that the multilevel dynamics are in the SH regime when $q > q_{WG}$ and the population should concentrate at the full-cooperator equilibrium.

In Figure \ref{fig:qdensities}, we illustrate the impact of introducing indirect reciprocity and increasing the defector detection probability $q$ on the steady-state densities of the multilevel dynamics for a fixed initial condition and relative strength of selection $\lambda$. The leftmost and lightest-colored curve corresponds to within-group PD interactions without detection and punishment of defectors ($q=0$), for which our value of $\lambda$ is only slightly over the threshold level $\lambda^*_{PD}$ needed to produce a steady state density supporting some cooperation. As we increase the parameter $q$, the steady-state densities with darker-colored curves show increased support for cooperation, with almost all groups consisting mostly of cooperators by the time $q = 0.625$. This is consistent with our result that the steady-state densities should concentrate upon a delta-function at full-cooperation when $q \to q_{WG}$, where $q_{WG} = 0.8$ for our choice of parameters.

 \begin{figure}[ht]
 	\centering
 	\includegraphics[width=0.65\textwidth
 	]{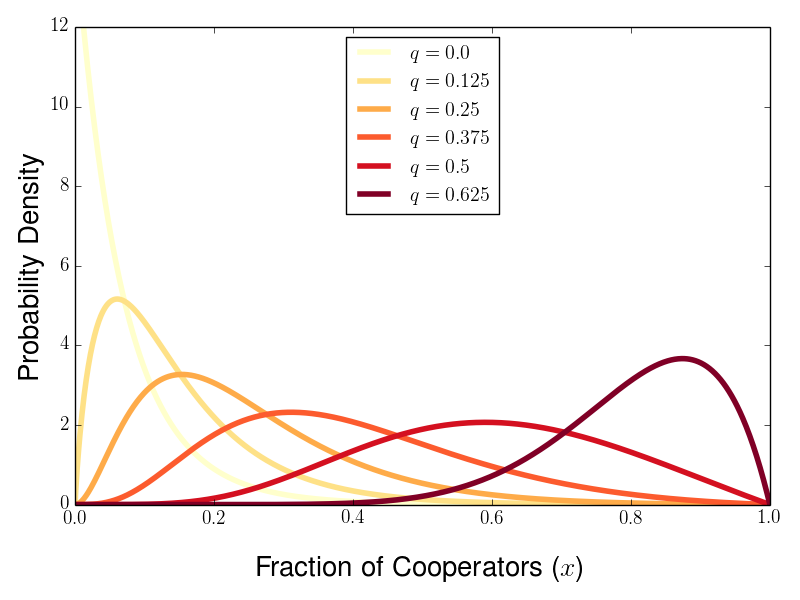}
	\caption[Steady state densities for fixed game and $\lambda$ with various values of the defector detection probability $q$,]{Steady state densities for $\lambda = 10$, $\gamma = 1.5$, $\theta = 2$, $\alpha = \beta = -1$ and various values of the defector detection probability $q$. For this choice of parameters, $q_{WG} = \tfrac{4}{5}$ and $q_{BG} = \tfrac{1}{3}$. }
	\label{fig:qdensities}
 \end{figure}

\subsection{Multilevel Selection in a Repeated Game} \label{sec:directreciprocity}

In this section, we will consider our multilevel dynamics when game-theoretic interactions correspond to a repeated Prisoners' Dilemma. Individuals will be paired up for social interactions and initially play a PD game with the payoff matrix from Equation \eqref{eq:payoffmatrix}. After each game is played, individuals continue their interaction by playing an additional game with probably $\delta$, and terminate their interaction with probability $1 - \delta$. As a result, a pair of individuals will a sequence of PD games against each other, and the expected length of their interaction is $\tfrac{1}{1-\delta}$ games. We consider two strategies: always defect (All-D), in which individuals defect in each round, and tit-for-tat (TFT), in which individuals cooperate in the round of an interaction and then copy the action of their opponent in each subsequent round. For consistency with our terminology from previous sections, we will often denote the TFT strategy by $C$ and the All-D strategy by $D$ when describing the individual-level and group-level replication rates. 

We now formulate the expected payoffs received by individuals following the TFT and All-D strategies. When two TFT players meet, they both start with cooperation and respond in each subsequent round by cooperation, receiving a reward $R$ for cooperation in each round and accruing an expected payoff of $R \left( \tfrac{1}{1-\delta} \right)$. When two All-D players meet, they will both defect and receive a punishment $P$ in each round, accruing an expected payoff of $P \left( \frac{1}{1-\delta}\right)$. Finally, when a TFT and All-D player meet, the TFT player cooperates in the first round, receiving payoff $S$, then defects and receives payoff $P$ in all subsequent rounds,  resullting in an expected payoff of $S + P \left( \tfrac{\delta}{1-\delta} \right)$. Correspondingly, an all-D player will receive an expected payoff of $T + P  \left( \tfrac{\delta}{1-\delta} \right)$ when paired with a TFT-player. 

We then assume that individuals will play the repeated PD against all of the members of their group. In a group composed of fraction $x$ TFT-players and $1-x$ fraction All-D players, we assume that individuals will play the expected payoff achieved by TFT-players and All-D players is given by
\begin{subequations} \label{eq:pidelta}
\begin{alignat}{2}
\pi^{\delta}_C(x) &= x \frac{1}{1-\delta} R + (1-x) \left( S +  \frac{\delta}{1 - \delta}  P\right) &&= \pi_C(x) + \frac{\delta}{1 - \delta} \left( xR + (1-x) P \right)\\
\pi^{\delta}_D(x)  &= x \left( T + \frac{1}{1-\delta} P \right) + (1-x) \frac{1}{1-\delta} P && = \pi_D(x) + \frac{\delta}{1-\delta} P .
\end{alignat}
\end{subequations}

We can also describe this payoff achieved 
\begin{equation} \label{eq:deltapayoffmatrix}
\begin{blockarray}{ccc}
& \textnormal{TFT} & \textnormal{All-D} \\
\begin{block}{c(cc)}
\textnormal{TFT} & \ds\frac{R}{1-\delta}  & S + \ds\frac{\delta P}{1-\delta}   \\
\textnormal{All-D} & T + \ds\frac{\delta P}{1 - \delta}  & \ds\frac{P}{1 - \delta}  \\
\end{block}
\end{blockarray}
\end{equation}

Therefore the payoff difference between TFT players and All-D players in an $x$ cooperator group is 
\begin{equation} \label{eq:deltaprocesspayoffdiff} \pi_C^{\delta}(x) - \pi_D^{\delta}(x) = \pi_C(x) - \pi_D(x) + \frac{\delta}{1 - \delta} \left(R - P \right) x = \beta + \left[ \left(\frac{1}{1-\delta}\right) \alpha + \left(\frac{\delta}{1-\delta} \right) \gamma \right] x. \end{equation}
By plugging in $x = 0$, we see that $\pi_C^{\delta}(0) - \pi_D{\delta}(0) = \beta < 0$ for any PD game and for any $\delta < 0$, so the all-defector equilibrium is always stable under the within-group dynamics. For the all-cooperator equilibrium, we find that
\begin{equation}\label{eq:deltaprocessdefect}
  \pi_D^{\delta}(1) - \pi_C^{\delta}(1) = - \beta - \left[\left(\tfrac{1}{1-\delta} \right) \alpha +  \left(\tfrac{\delta}{1-\delta}  \right) \gamma \right],
\end{equation}
so the all-cooperator equilibrium becomes locally stable when the continuation probability $\delta$ satisfies
 \begin{equation} \label{eq:deltaW} \delta > \delta_{W} := \frac{-(\alpha + \beta)}{\gamma - \beta}. \end{equation} 
We can also use Equation \ref{eq:pidelta} to see that the average payoff of group members is given by
\begin{equation} \label{eq:Gdelta} G_{\delta}(x) = P + \gamma x + \left(\alpha + \frac{\delta}{1-\delta} \left( \gamma + \alpha \right) \right) x^2, \end{equation}
and that the average payoff in a full-cooperator group is 
\begin{equation} \label{eq:deltaGfull}  G_{\delta}(1) = \gamma + \alpha + \left( \frac{\delta}{1-\delta} \right) \left(\gamma + \alpha \right) = \left( \frac{1}{1-\delta} \right) \left( \gamma + \alpha\right). \end{equation}
Using the quantities calculated above for individual and collective payoff functions $\pi_C^{\delta}(x)$, $\pi_D^{\delta}(x)$, and $G_{\delta}(x)$, we see the bifurcation picture and general possibilities for long-time behavior in our model for multilevel selection with direct reciprocity are analogous to that of the indirect reciprocity model presented in Figure \ref{fig:qbifurcation} and Table \ref{tab:qtable}.

When $0 \leq \delta < \delta_W$, the multilevel dynamics are in the PD regime, so we expect to see density steady states in the form of Equation \ref{eq:PDsteady} when the relative of between-group selection $\lambda$ is sufficiently large. In particular, we can use Equation \ref{eq:PDthresh} to see that the threshold to establish cooperation through multilevel selection is given by
\begin{equation} \label{eq:deltathresh} \lambda^*_{\delta} = \frac{\theta}{\gamma + \alpha} \left[  \beta + \left( \frac{1}{1-\delta} \right) \alpha + \left(\frac{\delta}{1 - \delta} \right) \gamma \right], \end{equation}
which is a decreasing function of the continuation probability $\delta$. Similarly, we can apply our formulas for direct reciprocity to Equation \eqref{eq:steadystatefitness} to see that, when $0 \leq \delta < \delta_W$, the average payoff at the steady state density 
\begin{equation} \label{eq:deltaGav} \langle G^{\delta}(\cdot) \rangle_{f(x)} = \frac{ \gamma + \alpha}{1- \delta} + \frac{1}{\lambda} \left[ \beta + \left( \frac{1}{1-\delta} \right) \alpha + \left(\frac{\delta}{1 - \delta} \right) \gamma \right], \end{equation}
 which is increasing in $\delta$. Furthermore, in the limit as $\lambda \to \infty$, the long-time average payoff achieved by the population is given by
 \begin{equation}
 \lim_{\lambda \to \infty} \langle G^{\delta}(\cdot) \rangle_{f^{\lambda}_{\theta}}  = \frac{\gamma + \alpha}{1 - \delta},
 \end{equation}
which, for $\delta > 0$, exceeds the collective payoff achieved by an all-cooperator group under the original game of Equation \eqref{eq:payoffmatrix} under well-mixed one-shot interactions. 

In Figure \ref{fig:deltaunnorm}, we illustrate the average payoff of the population at steady state for various values of relative selection strength $\lambda$ and continuation probability $\delta$. Unlike the previously studied mechanisms, we see that higher values of average payoff are achieved for larger continuation probabilities than for the average payoff achieved in the limit of strong between-group competition ($\lambda \to \infty$) under the multilevel dynamics for the case of well-mixed, one-shot interactions. We also illustrate the threshold level of relative selection strength $\lambda^*_{\delta}$ needed to sustain cooperation at steady state by the gray dashed line.

\begin{figure}[ht]
    \centering

    \hspace{-7mm} \includegraphics[width=0.6\textwidth
    ]{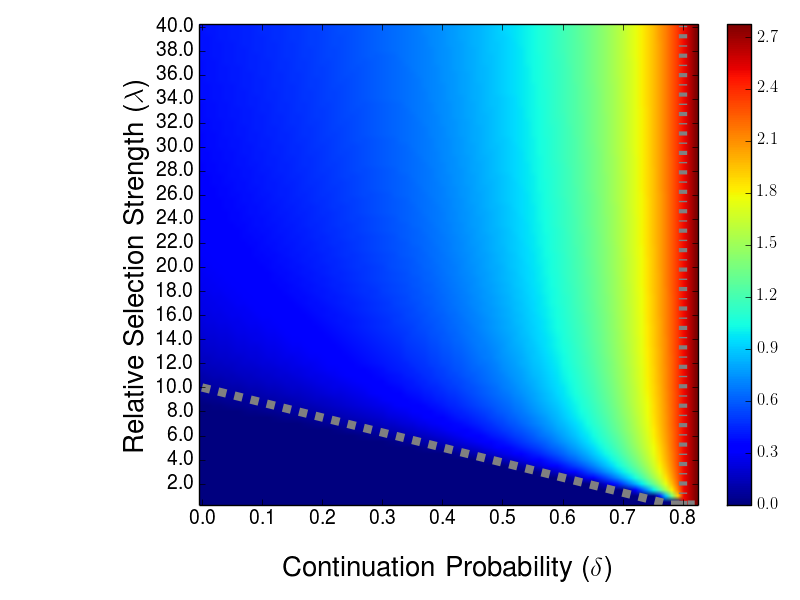}

    \caption[Average payoff of the population at steady state for various values of $\lambda$ and $\delta$ in the multilevel repeated PD dynamics.]{Average payoff of the population at steady state for various values of $\lambda$ and $\delta$ in the multilevel repeated PD dynamics. Game-theoretical parameters are given by $\alpha = \beta = -1$ and $\gamma = 1.5$, and then initial condition has H{\"o}lder exponent $\theta = 2$ near the all-cooperator equilibrium. The downward sloping dotted gray line represents the threshold level of $\lambda^*_{\delta}$ need to produce cooperation for given continuation probabilities $\delta$. The vertical dash-dotted gray line represents the continutation probability above which the all-cooperator equilibrium becomes locally stable $\delta_W$, which is equal to $0.8$ for our choice of parameters. }
    \label{fig:deltaunnorm}
\end{figure}

As we see that allowing repeated interactions improves the maximal possible payoff achieved by groups, it is potentially unfair to measure payoff for an entire repeated game in terms of the payoff matrix of the one-shot stage game. Alternatively, we can consider the possibility of measuring the payoff of the repeated game in the units of the stage game, using the notion of discounted average payoff \cite{fudenberg1991game}. Under this approach, we multiply individual payoffs by $1-\delta$ to balance the denominator $\frac{1}{1-\delta}$ arising from the receipt of expected payoffs over the potentially infinite horizon of the game. Therefore the discounted average payoffs $\tilde{\pi}_C(x) := (1-\delta) \pi_C(x)$ and $\tilde{\pi}_{D}(x) := (1-\delta) \pi_D(x)$ for the TFT and All-D strategies are given by
\begin{subequations} \label{eq:pideltadiscounted}
\begin{align}
\pi^{\delta}_C(x) &= \left(1 - \delta \right)\pi_C(x) +\delta\left( \gamma + \alpha \right) + \delta P\\
\pi^{\delta}_D(x) &= \left(1 - \delta \right)\pi_D(x) + \delta P .
\end{align}
\end{subequations}
From these expressions, we see that the discounted average payoffs under direct reciprocity take the same form as the payoffs $\pi_C^q(x)$ and $\pi_D^q(x)$ from Equation \ref{eq:piqx} from our model of indirect reciprocity, except with the continuation probability $\delta$ taking the place of the defector detection probability $q$.  Intuitively, we can understand this equivalence between the payoffs from our models of indirect reciprocity and of direct reciprocity with discounted average payoffs because interactions between a defector and a cooperator/TFT-player generates two punishment payoffs with a given probability (respectively $q$ and $\delta$) and generates an $S$ payoff and a $T$ payoff with the complementary probability. Under discounted average payoff, we then expect the dynamics of multilevel selection with direct reciprocity to agree with the results from Section \ref{sec:indirectreciprocity} under indirect reciprocity.  %

\section{Discussion} \label{sec:discussion}

In this paper, we considered the dynamics of a PDE model for multilevel selection arising from evolutionary game theory, exploring the effects of modifying within-group competition through the mechanisms of assortment, other-regarding preferences, and both direct and indirect reciprocity. We found that all four of these mechanisms helped to promote cooperation via multilevel selection, improving the outcome from multilevel competition relative to the case of well-mixed within-group interactions. Applying recent results for the long-time behavior for this class of PDE models, we were able to characterize the effects of the within-group mechanisms on both the threshold relative strength of between-group selection needed to sustain cooperation at steady state and on the average payoff achieved in a population at steady state. For each mechanism, we find that there is a regime in which the mechanism is sufficiently weak that it cannot help promote any long-time cooperation under individual-level selection, but that the mechanism can help to faciliate the survival of cooperation in concert with multilevel selection. 

We illustrate this synergistic effect in Figure \ref{fig:rsynergyplot}, showing, in the case of the assortment model, that there is a broad region of the parameter space of between-group selection strength $\lambda$ and the assortment probability $r$ such that cooperation can be achieved through a combination of assortment and multilevel selection, while defection would dominate in the absence of either of the two mechanisms. A similar figure can be produced to illustrate the synergy of within-group population structure and between-group competition for the promotion of cooperation, highlighting how population structure can help to decrease the individual-level incentive to defect and thereby facilitate the collective achievement of greater levels of cooperation.

 \begin{figure}[ht]
 	\centering
 	\includegraphics[width=0.6\textwidth]{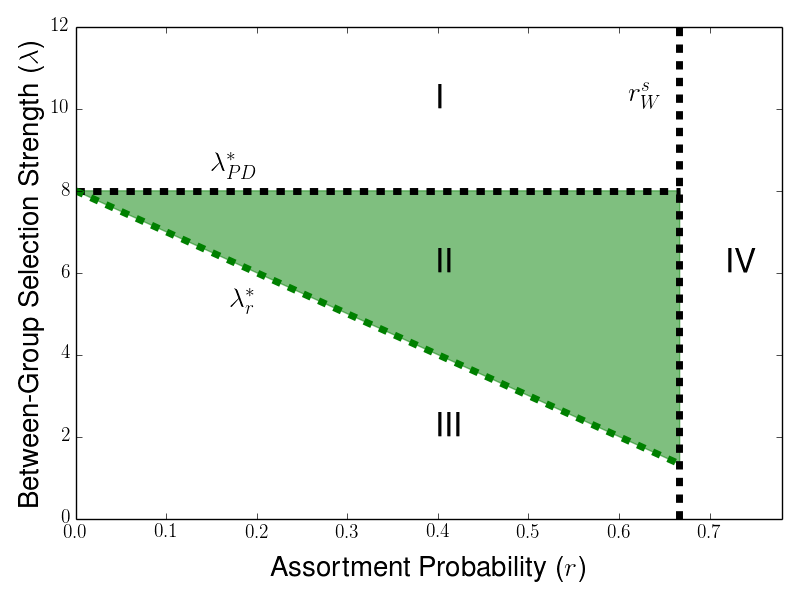}
	\caption{Illustration of regions of parameter space of between-group selection strength $\lambda$ and assortment probability $r$ in which cooperation can be supported under individual-level selection with assortative interactions, under multilevel selection, or with a combination of the two mechanisms. Given a fixed initial distribution of strategies, we can divide the parameter space into the following four regions: in which cooperation survives under multilevel selection with well-mixed interactions (Region I), in which cooperation can only survive under the combination of assortment and multilevel selection (Region II, highlighted in green), in which defection dominates the population even in the presence of assortment and multilevel selection (Region III), and in which cooperation can survive by individual-level selection alone in the presence of assortative interactions (Region IV). }
	\label{fig:rsynergyplot}
 \end{figure}

While we have seen that each of the mechanisms considered helps to facilitate long-time cooperation and increase the average payoff at steady state for finite strengths of between-group selection, we see that the mechanism differed in their ability to improve the maximal collective outcome achievable under the dynamics of multilevel selection. For our models of assortment and indirect reciprocity, the collective outcome achieved in the limit of infinite strength of between-group competition ($\lambda \to \infty$) was still the payoff of the all-cooperator group ($G(1)$), even under fully assortative interactions ($r = 1$) or when cooperators always punished defectors ($q =1$). As a result, these two within-group mechanisms were not able to erase the shadow cast by lower-level selection, as, for PD games in which an intermediate level of cooperation maximized collective payoff under well-mixed interactions, no level of assortment or reciprocity could allow the achievement of the socially optimal payoff.

 For our model of direct reciprocity, the maximal achievable average payoff was $G_{\delta}(1) = (1-\delta)^{-1} G(1)$, so the presence of repeated interactions did result in improved collective outcomes, with the caveat that repeated interactions naturally increases the total available payoff from interactions between individuals. Finally, for our model with other-regarding preferences, we saw that, when indvidual-level reproduction rates place sufficient weight on the payoffs of one's opponents we were able to see that the dynamics of indvidual-level and multilevel selection can promote collective outcomes that can outperform the all-cooperator group. Furthermore, the socially optimal collective outcome is always achieved in the limit when individual payoffs place equal weight on one's own payoff and the payoffs of one's opponents, achieving perfect alignment of individual and collective evolutionary interests.

From a modeling perspective, the difference between the maximal collective outcomes depends on the ways in which each mechanism alters the individual-level and group-level evolutionary incentives. Each of these mechanisms decreases the individual-level incentive to defect in a group with many cooperators, which is why we see the decrease in threshold between-group selection strength and the increase of average payoff at steady state when there is a marginal increase in the strength of each of our mechanisms (corresponding to increasing the parameters $r$, $F$, $q$, or $\delta$ in each of our models). However, our mechanisms differ in how they serve to either promote cooperation or help increase average payoff of the group. Our model of assortment clusters cooperators with cooperators and defectors with defectors, serving to increase the indvidual-level incentive to cooperate even when an intermediate level of cooperation may produce the best average outcome in a group. Similarly, both the models of direct and indirect reciprocity considered in this paper serve to decrease the individual incentive to defect, and results in punishing defectors even when some presence of defectors would actually benefit the group. By contrast, the mechanism of other-regarding preference serves to incentivize individuals to care about the impacts of their actions on their opponents, connecting the individual benefit derived from a game-theoretic interaction with the contribution that this interaction makes towards the average payoff of the group. As other-regarding preference is the only mechanism we consider that emphasizes improving average payoff rather than increasing cooperation per se, it makes sense that this is the only mechanism that can allow achievement of collective outcomes exceeding the all-cooperator payoff when the underlying game most favors a mix of cooperators and defectors under between-group replication. 

From a mathematical perspective, the analysis of the models in this paper highlights two different key routes for mechanisms to improve the best possible collective outcome achievable under the long-time dynamics of multilevel selection. The first option is for a mechanism to increase the collective reproduction rate $G(1)$ of the all-cooperator group, which occurs in our model of direct reciprocity. The second option is for a mechanism to change the within-group dynamics such that a new equilibrium $x_{eq}$ can feature a greater collective reproduction-rate $G(x_{eq})$ than that of the all-cooperator group. This can occur either through a bifurcation creating a stable within-group equilibrium, as seen in our model of other-regarding preferences, or by the creating of a new unstable equilibrium, as seen in a recent paper on a PDE model for protocell evolution through the introduction of a third strategy \cite{cooney2021pde}. The identification of these two routes to improving the best possible collective payoff can be applied in future work to understand how a range of within-group mechanisms can work in concert with multilevel selection to help promote cooperative behaviors, from homophilous processes like active linking \cite{pacheco2006active} and assortative matching \cite{bergstrom2003algebra,bergstrom1995evolution,bergstrom2013measures} to reciprocity mechanisms like social norms \cite{axelrod1986evolutionary,nowak2005indirect,ohtsuki2006leading,pacheco2006stern} and the social ostracism of defectors \cite{tavoni2012survival,tilman2017maintaining}.

In prior work by Nowak and coauthors comparing the achievement of cooperation under multilevel selection, assortment, and reciprocity, these mechanisms were presented alongside models in which game-theoretic interactions took place on $k$-regular graphs. Introduced by Ohtuski and coauthors, these models for evolutionary games on graphs showed how a group of individuals having game-theoretic interactions and competing for replication with neighbors on a graph could achieve cooperation for games in which defection dominated under well-mixed interactions \cite{ohtsuki2006simple,ohtsuki2006replicator}. Ohtsuki and Nowak used a pair-approximation to derive a replicator equation for individual-level selection on $k$-regular graphs for a variety of update rules, and characterize the resulting dynamics in terms of a transformed payoff matrix \cite{ohtsuki2006replicator}. Unlike the transformed payoff matrices for the models discussed in this paper, the Ohtuski-Nowak payoff transformation does not actually describe the payoffs achieved by cooperators and defectors with interactions taking place on a graph, and therefore this payoff matrix cannot be used to formulate the function $G(x)$ describing the average payoffs of group members in an $x$-cooperator group. To study multilevel selection for the case in which within-group interactions and replication competition take place on a $k$-regular graph, we will need to return to the pair-approximation to derive an appropriate description of between-group competition and the resulting PDE model for multilevel competition. The details of this analysis are carried out in a forthcoming paper, and it is shown how the number of graph neighbors $k$ and the update rule for individual-level selection (death-birth, birth-death, or imitation) impact the level of cooperation achieved via multilevel selection.

 Multilevel selection has also been suggested to work in concert with the mechanisms of strong reciprocity via altruistic punishment \cite{gintis2000strong,boyd2003evolution,janssen2014effect}, institutional incentives for the mangament of common-pool resources \cite{waring2017coevolution}, and social norms for evaluating and incentivizing individuals based on social reputations \cite{santos2007multi,scheuring2010coevolution}. Simulation studies of cultural group selection have shown that the presence of altruistic punishers can result in greater levels of long-time cooperation than would be achieved under multilevel competition between defectors and non-punishing cooperators alone. Because the approach taken in this paper can be extended to study a broad family of within-group and between-group replication rates for multilevel competition between pairs of types \cite{cooney2021long}, a potential direction for future research could be to formulate and analyze analytically tractable versions of these models from cultural group selection for the evolution of altruistic punishment or the evolution of social norms. Such an approach would provide a useful baseline for further study in the dynamics of multilevel selection in cultural evolution, and to understand how the ability for social and cultural institutions to faciliate long-time cooperation may depend on the impact they have on the individual incentive to defect and the collective incentive to feature full-cooperation over full-defection. Furthermore, because many of these existing models study the multilevel competition with three players (defectors, cooperators, and either conditional cooperators or altruistic punishers), these mechanisms also motivate further numerical and analytical investigations for PDE models of multilevel selection with three types of individuals. By considering the different ways in which modifying interactions between individuals can impact the dynamics of PDE models of multilevel selection, we can both learn more about these mechanisms and identify new mathematical approaches for studying the class of non-local, hyperbolic PDEs that arise as replicator equations for selection at multiple levels. 

\renewcommand{\abstractname}{Acknowledgments}
\begin{abstract} 
 This research was supported by NSF through grants DMS-1514606 and GEO-1211972, by the ARO grant W911NF-18-1-0325, and by the Simons Foundation through the Math + X grant awarded to University of Pennsylvania. I am thankful to Simon Levin, Joshua Plotkin, Yoichiro Mori, Carl Veller, Chai Molina, Feng Fu, and Alex McAvoy for helpful discussions, and I would like to thank Joshua Plotkin and two anonymous referees for helpful comments on the manuscript.

\end{abstract}

\renewcommand{\abstractname}{Statement on Code Availability}
\begin{abstract} 
All code used to generate figures is archived on Github (\href{https://github.com/dbcooney/Multilevel-Mechanism-Paper-Code}{https://github.com/dbcooney/Multilevel-Mechanism-Paper-Code}) and licensed for reuse, with appropriate
attribution/citation, under a BSD 3-Clause Revised License.
\end{abstract}

\bibliographystyle{unsrt}
\bibliography{multilevelselection}

\begin{thebibliography}{10}

\bibitem{hogeweg2003multilevel}
Paulien Hogeweg and Nobuto Takeuchi.
\newblock Multilevel selection in models of prebiotic evolution: compartments
  and spatial self-organization.
\newblock {\em Origins of Life and Evolution of Biospheres}, 33(4):375--403,
  2003.

\bibitem{szathmary1987group}
E{\"o}rs Szathm{\'a}ry and L{\'a}szl{\'o} Demeter.
\newblock Group selection of early replicators and the origin of life.
\newblock {\em Journal of Theoretical Biology}, 128(4):463--486, 1987.

\bibitem{szathmary1995major}
E{\"o}rs Szathm{\'a}ry and John~Maynard Smith.
\newblock The major evolutionary transitions.
\newblock {\em Nature}, 374(6519):227--232, 1995.

\bibitem{boza2010beneficial}
Gergely Boza and Szabolcs Sz{\'a}mad{\'o}.
\newblock Beneficial laggards: multilevel selection, cooperative polymorphism
  and division of labour in threshold public good games.
\newblock {\em BMC Evolutionary Biology}, 10(1):336, 2010.

\bibitem{shaffer2016foundress}
Zachary Shaffer, Takao Sasaki, Brian Haney, Marco Janssen, Stephen~C Pratt, and
  Jennifer~H Fewell.
\newblock The foundress’s dilemma: group selection for cooperation among
  queens of the harvester ant, {Pogonomyrmex californicus}.
\newblock {\em Scientific Reports}, 6:29828, 2016.

\bibitem{van2019role}
Simon Van~Vliet and Michael Doebeli.
\newblock The role of multilevel selection in host microbiome evolution.
\newblock {\em Proceedings of the National Academy of Sciences},
  116(41):20591--20597, 2019.

\bibitem{gilchrist2004optimizing}
Michael~A Gilchrist, Daniel Coombs, and Alan~S Perelson.
\newblock Optimizing within-host viral fitness: infected cell lifespan and
  virion production rate.
\newblock {\em Journal of Theoretical Biology}, 229(2):281--288, 2004.

\bibitem{levin1981selection}
Simon Levin and David Pimentel.
\newblock Selection of intermediate rates of increase in parasite-host systems.
\newblock {\em The American Naturalist}, 117(3):308--315, 1981.

\bibitem{blackstone2020variation}
Neil~W Blackstone, Sarah~R Blackstone, and Anne~T Berg.
\newblock Variation and multilevel selection of {SARS-CoV-2}.
\newblock {\em Evolution}, 74(10):2429--2434, 2020.

\bibitem{traulsen2005stochastic}
Arne Traulsen, Anirvan~M Sengupta, and Martin~A Nowak.
\newblock Stochastic evolutionary dynamics on two levels.
\newblock {\em Journal of Theoretical Biology}, 235(3):393--401, 2005.

\bibitem{traulsen2006evolution}
Arne Traulsen and Martin~A Nowak.
\newblock Evolution of cooperation by multilevel selection.
\newblock {\em Proceedings of the National Academy of Sciences},
  103(29):10952--10955, 2006.

\bibitem{traulsen2008analytical}
Arne Traulsen, Noam Shoresh, and Martin~A Nowak.
\newblock Analytical results for individual and group selection of any
  intensity.
\newblock {\em Bulletin of Mathematical Biology}, 70(5):1410, 2008.

\bibitem{simon2010dynamical}
Burton Simon.
\newblock A dynamical model of two-level selection.
\newblock {\em Evolutionary Ecology Research}, 12(5):555--588, 2010.

\bibitem{markvoort2014computer}
Albert~J Markvoort, Sam Sinai, and Martin~A Nowak.
\newblock Computer simulations of cellular group selection reveal mechanism for
  sustaining cooperation.
\newblock {\em Journal of Theoretical Biology}, 357:123--133, 2014.

\bibitem{bottcher2016promotion}
Marvin~A B{\"o}ttcher and Jan Nagler.
\newblock Promotion of cooperation by selective group extinction.
\newblock {\em New Journal of Physics}, 18(6):063008, 2016.

\bibitem{grafen1979hawk}
Alan Grafen.
\newblock The hawk-dove game played between relatives.
\newblock {\em Animal Behaviour}, 27(4):905--907, 1979.

\bibitem{eshel1982assortment}
Ilan Eshel and Luigi~Luca Cavalli-Sforza.
\newblock Assortment of encounters and evolution of cooperativeness.
\newblock {\em Proceedings of the National Academy of Sciences},
  79(4):1331--1335, 1982.

\bibitem{smith1982evolution}
John Maynard~Smith.
\newblock {\em Evolution and the Theory of Games}.
\newblock Cambridge University Press, 1982.

\bibitem{trivers1971evolution}
Robert~L Trivers.
\newblock The evolution of reciprocal altruism.
\newblock {\em The Quarterly Review of Biology}, 46(1):35--57, 1971.

\bibitem{nowak2005indirect}
Martin~A Nowak and Karl Sigmund.
\newblock Evolution of indirect reciprocity.
\newblock {\em Nature}, 437(7063):1291, 2005.

\bibitem{ohtsuki2006leading}
Hisashi Ohtsuki and Yoh Iwasa.
\newblock The leading eight: social norms that can maintain cooperation by
  indirect reciprocity.
\newblock {\em Journal of theoretical biology}, 239(4):435--444, 2006.

\bibitem{durrett1994importance}
Richard Durrett and Simon Levin.
\newblock The importance of being discrete (and spatial).
\newblock {\em Theoretical Population Biology}, 46(3):363--394, 1994.

\bibitem{killingback1996spatial}
Timothy Killingback and Michael Doebeli.
\newblock Spatial evolutionary game theory: Hawks and doves revisited.
\newblock {\em Proceedings of the Royal Society of London. Series B: Biological
  Sciences}, 263(1374):1135--1144, 1996.

\bibitem{ohtsuki2006simple}
Hisashi Ohtsuki, Christoph Hauert, Erez Lieberman, and Martin~A Nowak.
\newblock A simple rule for the evolution of cooperation on graphs and social
  networks.
\newblock {\em Nature}, 441(7092):502--505, 2006.

\bibitem{ohtsuki2006replicator}
Hisashi Ohtsuki and Martin~A Nowak.
\newblock The replicator equation on graphs.
\newblock {\em Journal of Theoretical Biology}, 243(1):86--97, 2006.

\bibitem{nowak2006five}
Martin~A Nowak.
\newblock Five rules for the evolution of cooperation.
\newblock {\em Science}, 314(5805):1560--1563, 2006.

\bibitem{taylor2007transforming}
Christine Taylor and Martin~A Nowak.
\newblock Transforming the dilemma.
\newblock {\em Evolution: International Journal of Organic Evolution},
  61(10):2281--2292, 2007.

\bibitem{luo2014unifying}
Shishi Luo.
\newblock A unifying framework reveals key properties of multilevel selection.
\newblock {\em Journal of Theoretical Biology}, 341:41--52, 2014.

\bibitem{van2014simple}
Matthijs van Veelen, Shishi Luo, and Burton Simon.
\newblock A simple model of group selection that cannot be analyzed with
  inclusive fitness.
\newblock {\em Journal of Theoretical Biology}, 360:279--289, 2014.

\bibitem{luo2017scaling}
Shishi Luo and Jonathan~C Mattingly.
\newblock Scaling limits of a model for selection at two scales.
\newblock {\em Nonlinearity}, 30(4):1682, 2017.

\bibitem{cooney2019replicator}
Daniel~B Cooney.
\newblock The replicator dynamics for multilevel selection in evolutionary
  games.
\newblock {\em Journal of Mathematical Biology}, 79(1):101--154, 2019.

\bibitem{cooney2019analysis}
Daniel~B Cooney.
\newblock Analysis of multilevel replicator dynamics for general two-strategy
  social dilemma.
\newblock {\em Bulletin of Mathematical Biology}, 82(6):66, 2020.

\bibitem{cooney2021long}
Daniel~B Cooney and Yoichiro Mori.
\newblock Long-time behavior of a pde replicator equation for multilevel
  selection in group-structured populations.
\newblock {\em arXiv preprint arXiv:2104.00392}, 2021.

\bibitem{simon2012numerical}
Burton Simon and Aaron Nielsen.
\newblock Numerical solutions and animations of group selection dynamics.
\newblock {\em Evolutionary Ecology Research}, 14(6):757--768, 2012.

\bibitem{simon2013towards}
Burton Simon, Jeffrey~A Fletcher, and Michael Doebeli.
\newblock Towards a general theory of group selection.
\newblock {\em Evolution}, 67(6):1561--1572, 2013.

\bibitem{simon2016group}
Burton Simon and Michael Pilosov.
\newblock Group-level events are catalysts in the evolution of cooperation.
\newblock {\em Journal of Theoretical Biology}, 410:125--136, 2016.

\bibitem{puhalskii2017large}
A~Puhalskii, M~Reiman, and B~Simon.
\newblock A large-population limit for a markovian model of group-structured
  populations.
\newblock {\em arXiv preprint arXiv:1712.09119}, 2017.

\bibitem{boyd2003evolution}
Robert Boyd, Herbert Gintis, Samuel Bowles, and Peter~J Richerson.
\newblock The evolution of altruistic punishment.
\newblock {\em Proceedings of the National Academy of Sciences},
  100(6):3531--3535, 2003.

\bibitem{santos2007multi}
Francisco~C Santos, Fabio~ACC Chalub, and Jorge~M Pacheco.
\newblock A multi-level selection model for the emergence of social norms.
\newblock In {\em European Conference on Artificial Life}, pages 525--534.
  Springer, 2007.

\bibitem{janssen2014effect}
Marco~A Janssen, Miles Manning, and Oyita Udiani.
\newblock The effect of social preferences on the evolution of cooperation in
  public good games.
\newblock {\em Advances in Complex Systems}, 17(03n04):1450015, 2014.

\bibitem{michod1996cooperation}
Richard~E Michod.
\newblock Cooperation and conflict in the evolution of individuality. ii.
  conflict mediation.
\newblock {\em Proceedings of the Royal Society of London. Series B: Biological
  Sciences}, 263(1372):813--822, 1996.

\bibitem{gabriel1960primitive}
Mordecai~L Gabriel.
\newblock Primitive genetic mechanisms and the origin of chromosomes.
\newblock {\em The American Naturalist}, 94(877):257--269, 1960.

\bibitem{smith1993origin}
J~Maynard Smith and Eors Sz{\'a}thmary.
\newblock The origin of chromosomes i. selection for linkage.
\newblock {\em Journal of Theoretical Biology}, 164(4):437--446, 1993.

\bibitem{szathmary1993evolution}
E~Szathm{\'a}ry and J~Maynard Smith.
\newblock The evolution of chromosomes ii. molecular mechanisms.
\newblock {\em Journal of theoretical biology}, 164(4):447--454, 1993.

\bibitem{cooney2021pde}
Daniel~B Cooney, Fernando~W Rossine, Dylan~H Morris, and Simon~A Levin.
\newblock A pde model for protocell evolution and the origin of chromosomes via
  multilevel selection.
\newblock {\em arXiv preprint arXiv:2109.09357}, 2021.

\bibitem{nowak2006evolutionary}
Martin~A Nowak.
\newblock {\em Evolutionary dynamics}.
\newblock Harvard University Press, 2006.

\bibitem{strauss2007partial}
Walter~A Strauss.
\newblock {\em Partial differential equations: An introduction}.
\newblock John Wiley \& Sons, 2007.

\bibitem{evans1998partial}
L.C. Evans and American~Mathematical Society.
\newblock {\em Partial Differential Equations}.
\newblock Graduate Studies in Mathematics. American Mathematical Society, 1998.

\bibitem{taylor1978evolutionary}
Peter~D Taylor and Leo~B Jonker.
\newblock Evolutionary stable strategies and game dynamics.
\newblock {\em Mathematical Biosciences}, 40(1-2):145--156, 1978.

\bibitem{hofbauer1998evolutionary}
Josef Hofbauer and Karl Sigmund.
\newblock {\em Evolutionary games and population dynamics}.
\newblock Cambridge University Press, 1998.

\bibitem{sandholm2010population}
William~H Sandholm.
\newblock {\em Population games and evolutionary dynamics}.
\newblock MIT press, 2010.

\bibitem{cressman2014replicator}
Ross Cressman and Yi~Tao.
\newblock The replicator equation and other game dynamics.
\newblock {\em Proceedings of the National Academy of Sciences}, 111(Supplement
  3):10810--10817, 2014.

\bibitem{kaznatcheev2018effective}
Artem Kaznatcheev.
\newblock Effective games and the confusion over spatial structure.
\newblock {\em Proceedings of the National Academy of Sciences},
  115(8):E1709--E1709, 2018.

\bibitem{van2017hamilton}
Matthijs van Veelen, Benjamin Allen, Moshe Hoffman, Burton Simon, and Carl
  Veller.
\newblock Hamilton's rule.
\newblock {\em Journal of Theoretical Biology}, 414:176--230, 2017.

\bibitem{allen2015games}
Benjamin Allen and Martin~A Nowak.
\newblock Games among relatives revisited.
\newblock {\em Journal of Theoretical Biology}, 378:103--116, 2015.

\bibitem{cooney2016assortment}
Daniel Cooney, Benjamin Allen, and Carl Veller.
\newblock Assortment and the evolution of cooperation in a {Moran} process with
  exponential fitness.
\newblock {\em Journal of Theoretical Biology}, 409:38--46, 2016.

\bibitem{coder2018effects}
Kira Coder~Gylling and {\AA}ke Br{\"a}nnstr{\"o}m.
\newblock Effects of relatedness on the evolution of cooperation in nonlinear
  public goods games.
\newblock {\em Games}, 9(4):87, 2018.

\bibitem{iyer2020evolution}
Swami Iyer and Timothy Killingback.
\newblock Evolution of cooperation in social dilemmas with assortative
  interactions.
\newblock {\em Games}, 11(4):41, 2020.

\bibitem{van2011replicator}
Matthijs Van~Veelen.
\newblock The replicator dynamics with n players and population structure.
\newblock {\em Journal of Theoretical Biology}, 276(1):78--85, 2011.

\bibitem{szabo2012selfishness}
Gyoergy Szabo and Attila Szolnoki.
\newblock Selfishness, fraternity, and other-regarding preference in spatial
  evolutionary games.
\newblock {\em Journal of Theoretical biology}, 299:81--87, 2012.

\bibitem{pena2015evolutionary}
Jorge Pe{\~n}a, Georg N{\"o}ldeke, and Laurent Lehmann.
\newblock Evolutionary dynamics of collective action in spatially structured
  populations.
\newblock {\em Journal of Theoretical Biology}, 382:122--136, 2015.

\bibitem{nowak1998evolution}
Martin~A Nowak and Karl Sigmund.
\newblock Evolution of indirect reciprocity by image scoring.
\newblock {\em Nature}, 393(6685):573--577, 1998.

\bibitem{fudenberg1991game}
D.~Fudenberg and J.~Tirole.
\newblock {\em Game Theory}.
\newblock MIT Press. MIT Press, 1991.

\bibitem{pacheco2006active}
Jorge~M Pacheco, Arne Traulsen, and Martin~A Nowak.
\newblock Active linking in evolutionary games.
\newblock {\em Journal of Theoretical Biology}, 243(3):437--443, 2006.

\bibitem{bergstrom2003algebra}
Theodore~C Bergstrom.
\newblock The algebra of assortative encounters and the evolution of
  cooperation.
\newblock {\em International Game Theory Review}, 5(03):211--228, 2003.

\bibitem{bergstrom1995evolution}
Theodore~C Bergstrom.
\newblock On the evolution of altruistic ethical rules for siblings.
\newblock {\em The American Economic Review}, pages 58--81, 1995.

\bibitem{bergstrom2013measures}
Theodore~C Bergstrom.
\newblock Measures of assortativity.
\newblock {\em Biological Theory}, 8(2):133--141, 2013.

\bibitem{axelrod1986evolutionary}
Robert Axelrod.
\newblock An evolutionary approach to norms.
\newblock {\em American Political Science Review}, 80(4):1095--1111, 1986.

\bibitem{pacheco2006stern}
Jorge~M Pacheco, Francisco~C Santos, and Fabio AC~C Chalub.
\newblock Stern-judging: A simple, successful norm which promotes cooperation
  under indirect reciprocity.
\newblock {\em PLoS Computational Biology}, 2(12):e178, 2006.

\bibitem{tavoni2012survival}
Alessandro Tavoni, Maja Schl{\"u}ter, and Simon Levin.
\newblock The survival of the conformist: social pressure and renewable
  resource management.
\newblock {\em Journal of Theoretical Biology}, 299:152--161, 2012.

\bibitem{tilman2017maintaining}
Andrew~R Tilman, James~R Watson, and Simon Levin.
\newblock Maintaining cooperation in social-ecological systems.
\newblock {\em Theoretical Ecology}, 10(2):155--165, 2017.

\bibitem{gintis2000strong}
Herbert Gintis.
\newblock Strong reciprocity and human sociality.
\newblock {\em Journal of Theoretical Biology}, 206(2):169--179, 2000.

\bibitem{waring2017coevolution}
Timothy~M Waring, Sandra~H Goff, and Paul~E Smaldino.
\newblock The coevolution of economic institutions and sustainable consumption
  via cultural group selection.
\newblock {\em Ecological Economics}, 131:524--532, 2017.

\bibitem{scheuring2010coevolution}
Istv{\'a}n Scheuring.
\newblock Coevolution of honest signaling and cooperative norms by cultural
  group selection.
\newblock {\em BioSystems}, 101(2):79--87, 2010.

\end{thebibliography}

\end{document}